\documentclass[12pt]{article}
\usepackage{amsmath}
\usepackage{graphicx,psfrag,epsf}
\usepackage{enumerate}
\usepackage{natbib}

\newcommand{\blind}{0}

\usepackage{lscape} 

\usepackage{arydshln}
\usepackage{multirow}
\usepackage{amsmath}
\usepackage{amssymb}
\usepackage{arydshln}
\usepackage{amsfonts}
\usepackage{algorithm}
\usepackage{algorithmic}
\usepackage{caption}
\usepackage{subfigure}
\usepackage{subfloat}
\usepackage{ntheorem}
\usepackage{enumerate}
\usepackage{verbatim}
\usepackage{graphicx}
\newtheorem{theorem}{Theorem}

\newtheorem{proposition}{Proposition}
\newtheorem{definition}{Definition}
\newtheorem{assumption}{Assumption}

\usepackage[colorlinks=true, allcolors=blue]{hyperref}
\usepackage{color}
\usepackage{multirow}
\definecolor{grey}{RGB}{80,96,150}

\newcommand{\xbold}{\boldsymbol{x}}
\newcommand{\ybold}{\boldsymbol{y}}
\newcommand{\zbold}{{\boldsymbol{z}}}
\newcommand{\thetabold}{{\boldsymbol{\theta}}}
\newcommand{\eg}{\emph{e.g.}}
\newcommand{\ie}{\emph{i.e.}}

\begin{document}

\def\spacingset#1{\renewcommand{\baselinestretch}%
{#1}\small\normalsize} \spacingset{1}


\if0\blind
{
  \title{\bf An adaptive approximate Bayesian computation MCMC with Global-Local proposals }
  \author{Xuefei Cao$^1$, 
    Shijia Wang$^{2*}$ 
    and
    Yongdao Zhou$^1$\thanks{
    Address correspondence to: Dr.\ Shijia Wang ({\tt wangshj1@shanghaitech.edu.cn}) and Dr.\ Yongdao Zhou ({\tt ydzhou@nankai.edu.cn}).}\\
     $^1$NITFID, School of Statistics and Data Science, Nankai University, China\\
    $^2$Institute of Mathematical Sciences, ShanghaiTech University, China\\
    }
  \maketitle
} \fi

\if1\blind
{
  \bigskip
  \bigskip
  \bigskip
  \begin{center}
    {\LARGE\bf  }
\end{center}
  \medskip
} \fi

\bigskip
\begin{abstract}

In this paper, we address the challenge of Markov Chain Monte Carlo (MCMC) algorithms within the approximate Bayesian Computation (ABC) framework, which often get trapped in local optima due to their inherent local exploration mechanism.
We propose a novel Global-Local ABC-MCMC algorithm that combines the ``exploration" capabilities of global proposals with the ``exploitation" finesse of local proposals.  
We integrate iterated importance resampling into the likelihood-free framework to establish an effective global proposal distribution. For high-dimensional parameter spaces, we optimize the efficiency of the local sampler by leveraging Langevin dynamics and common random numbers.
Furthermore, we introduce two adaptive schemes to enhance the algorithmic performance. The first scheme divides the update target of the importance proposal into a sequence of intermediate target distributions that progressively approximate the ABC posterior, thereby gradually updating the importance proposal distribution during the iterations. The second adaptive scheme automatically selects the optimal mixture of global and local moves through sequential optimization, based on a relative version of the expected squared jumping distance (ESJD).
We theoretically and numerically demonstrate that our method is able to improve sampling efficiency and achieve more reliable convergence for complex posteriors. We develop a software package that is available at \url{https://github.com/caofff/GL-ABC-MCMC}.
\end{abstract}

\noindent%

\spacingset{1.45}
\section{Introduction}
\newcounter{xxx}
\label{sec:intro}
Traditional Bayesian inference typically relies on the assumption that the likelihood functions of statistical models are tractable. However, for many complex applications, the likelihood functions have no explicit expressions or are difficult to estimate numerically. 
\textit{Approximate Bayesian computation} (ABC) \citep{Pritchard1999Population, Beaumont2002ABC} is a likelihood-free inference method based on simulations, it only requires to simulate  synthetic data from the model without evaluating the likelihood function. Rejection ABC \citep{tavare1997inferring,Pritchard1999Population} is the simplest and most intuitive version of ABC algorithms, which repeatedly draws samples of parameter $\thetabold$ independently from prior $\pi$, simulates synthetic data $\xbold$ for each sample of $\thetabold$, and rejects the parameter $\thetabold$ if the discrepancy $\mathfrak{D}(\xbold,\ybold)$ between the observed data $\ybold$ and the simulated data $\xbold$ exceeds a pre-specified tolerance threshold $\varepsilon$.  In practice, it is inefficient to sample parameters from a prior distribution, especially for high dimensional cases. Most of samples drawn from the prior distribution would fall into the low posterior region, which leads to very high rejection rates. Then accepting one sample may require thousands or even millions of draws. 

\cite{Marjoram2003MCMC} introduce an ABC {Markov chain Monte Carlo} (MCMC) algorithm to approximate the posterior distribution. 
Proposal distributions are typically chosen to generate local moves that depend on the last state of the chain in order to guarantee an admissible acceptance rate. Theoretically, a small tolerance value $\varepsilon$ leads to good posterior approximation. However, 
designing an efficient proposal to explore the parameter space, particularly in the case of high dimensional multimodal posterior distributions, poses a significant challenge. \cite{Wegmann2009MCMC} enhanced the performance of ABC-MCMC by relaxing the tolerance, incorporating subsampling, and applying regression adjustments to the MCMC output.  
\cite{Sisson2007SMC, del2012adaptive} integrated ABC-MCMC algorithms into the SMC framework \citep{del2006sequential}. The sequence of intermediate target distributions is defined by ABC posteriors, with a series of tolerance parameters. \cite{del2012adaptive} adaptively determines this sequence by managing the divergence of particles. 
\cite{Clarte2020Gibbs, Likelihood-freeapproximateGibbssampling} propose Gibbs version of the approximate Bayesian computation approaches that focuses on lower-dimensional conditional distributions. 
\cite{wang2022approximate} incorporates a flexible machine learning classifier into ABC, avoiding the need for summary statistics, to enable automated comparison of real and simulated data.
{Hamiltonian Monte Carlo} (HMC) \citep{DUANE1987HMC, neal2011mcmc} employ Hamiltonian dynamics to propose candidate parameters, efficiently avoiding the random walk behavior that can impede the mixing of proposals in high-dimensional spaces. \cite{meeds2015HABC} explored the application of HMC to ABC and proposed using \textit{completely random numbers} (CRNs) to improve the stability of gradient estimates to improve Hamiltonian dynamic. 

However, for a high dimensional multimodal posterior distribution, all local samplers mix poorly because the exploration is inherently slow and switching between isolated modes can be extremely rare. Independent proposals can generate more global updates to transfer particles between modes, but they are difficult to design, but a poor independent sampler in MCMC means an inefficient random walk, which will cause the algorithm to converge slowly. For likelihood-based Bayesian inference, \cite{samsonov2022local-global,gabrie2022adaptive} combine the local and global proposals to improve the efficiency of MCMC. \cite{cabezas2024markovian} introduces an adaptive tempering mechanism in continuous normalizing flows to learn the complex target distribution. Additionally, normalizing flows \citep{rezende2015variational, papamakarios2021normalizing} are used to improve the exploration capabilities of MCMC independent proposals.

In this paper, we propose a novel approach that combines the advantages of global and local MCMC proposals for challenging posterior distributions within a likelihood-free inference framework. Our work is inspired by \cite{samsonov2022local-global}, but beyond the fundamental differences between likelihood-based inference and approximate Bayesian computation, we develop an alternative structural design for integrating global and local moves. Concretely, at each iteration, the algorithm probabilistically selects either a global or a local proposal such that the particle update is exclusively global or local. Moreover, we propose a sequential optimization approach based on a unit cost version of expected square jump distance (cESJD) criterion, which is then used to adaptively adjust the probability of choosing the global proposal at each iteration. This adaptive strategy enables a flexible combination of global and local kernels based on their relative performance, thereby ensuring improved overall algorithmic efficiency.

Specifically, we propose a likelihood-free iterated-sampling importance resampling (i-SIR) approach to construct an efficient global ABC-MCMC move. By extending the results of \cite{samsonov2022local-global}, 
we show that ABC-i-SIR algorithm can be interpreted as a systematic-scan two-stage Gibbs sampler, and prove the uniform geometric ergodicity of the ABC-i-SIR Markov kernel. Furthermore, we also prove the V-geometric ergodicity of the combination of ABC-i-SIR Markov kernel and local Markov kernel, and hence show that under some conditions, the mixing rate of the Global-Local Markov Kernel is better than that of the local Markov kernel.

To further enhance the efficiency of the global ABC-MCMC move, 
we introduces intermediate target distributions with a decreasing sequence of tolerance within a single Markov chain, progressively and stably improve the proposal distribution using the weighted samples generated by ABC-i-SIR. 
We adaptively reduce the tolerance thresholds by controlling the proportion of alive training samples. This makes the selection of the proposal distribution more robust.

For high-dimensional parameter spaces, we employ the Metropolis-adjusted Langevin algorithm (MALA) to construct local ABC-MCMC moves. A key challenge in ABC-MALA is the estimation of the log-likelihood gradient, as the likelihood function is intractable. We utilize common random numbers to improve the accuracy and stability of the gradient estimates.

The rest of paper is organized as follows. Section \ref{sec:independent} presents the framework of our Global-Local ABC-MCMC, the efficient global MCMC move ABC-i-SIR and the efficient local MCMC move ABC-MALA. Section \ref{sec:adaptive} introduces two adaptive schemes for improving algorithmic efficiency: the adaptive updating of importance proposals and the sequential optimization of hyperparameters. The theoretical properties are established in Section \ref{sec:prop}. Sections \ref{sec:simu} and \ref{sec:real} demonstrate the effectiveness of our method through synthetic  examples studies and two real model analysis. Section \ref{sec:conc} provides the conclusion, and all proofs of the theoretical results are deferred to the Appendix.

\section{An ABC-MCMC with global and local moves}
\label{sec:independent}
We consider a given observed data set $\ybold\in\mathbb{Y}$, which is realized from a parametric model $\left\{P_{\thetabold_0}:\thetabold_0\in\right.$ $\left.\Theta\subset\mathbb{R}^p\right\}$. Assume that for each $\thetabold\in\Theta$, $P_\thetabold$ admits a density $p(\cdot\mid\thetabold)$ that cannot be directly evaluated but can be sampled from. 
The ABC posterior can be achieved by augmenting the target posterior from $\pi_\varepsilon(\thetabold\mid\ybold)\propto p(\ybold\mid\thetabold)\pi(\thetabold)$ to 
\begin{equation*}
\pi_\varepsilon(\thetabold,\xbold\mid\ybold)=\frac{\pi(\boldsymbol{\theta}) p(\xbold\mid\thetabold)K_\varepsilon(\xbold,\ybold)}{\iint\pi(\boldsymbol{\theta}) p(\xbold\mid\thetabold)K_\varepsilon(\xbold,\ybold)d\xbold d\thetabold},
\end{equation*}
where $\xbold$ denotes synthetic data simulated from $p(\cdot\mid\thetabold)$, $K_\varepsilon(\cdot,\cdot)$ is a kernel function and the threshold $\varepsilon$ plays a role of  bandwidth. The marginal posterior distribution $\pi_{\varepsilon}(\thetabold\mid \ybold)$ can be evaluated by integrating out the auxiliary data set $\xbold$ from $\pi_\varepsilon(\thetabold,\xbold\mid\ybold)$.
The choice of threshold $\varepsilon$ is a trade-off between accuracy and computational speed. A smaller value of threshold can lead to more accurate ABC posterior, but with higher computational cost.
When the threshold value $\varepsilon$ approaches to $0$, $p_\varepsilon(\ybold\mid\thetabold)$ converges to the exact Bayesian posterior $\pi(\ybold\mid\thetabold)$. 

\subsection{An ABC-MCMC with global and local moves}
\label{sec:basic}
Standard ABC methods based on rejection sampling become infeasible in high-dimensional parameter spaces due to poor acceptance rates. ABC-MCMC offers a more efficient alternative by constructing a Markov chain targeting \(\pi_{\varepsilon}(\boldsymbol{\theta}, \xbold \mid \ybold)\). 
While local proposals \(q(\cdot \mid \boldsymbol{\theta})\) in ABC-MCMC facilitate efficient exploration near the current state, they often suffer from poor mixing for complex or multimodal posteriors, as the chain can become trapped in local modes.

To enhance the exploration capabilities of MCMC and facilitate more frequent transitions between distinct modes, it is necessary to incorporate more global updates. However, designing efficient global proposals is challenging due to the unknown posterior distribution, which is analogous to the difficulty of performing posterior inference itself. Furthermore, proposing each parameter globally and independently often fails to capture the dependencies and correlations among parameters, hindering effective exploration of the parameter space. To address these challenges, we propose a Global-Local ABC-MCMC framework that integrates both local and global proposals. This approach leverages the complementary advantages of each proposal type, striking an optimal balance that enables thorough exploration of local regions while simultaneously ensuring a more comprehensive coverage of the entire parameter space.
Specifically, at each iteration, the algorithm alternates between a global move with probability \(\gamma\) and a local move with probability \(1 - \gamma\).

The most common and straightforward local move in ABC-MCMC involves generating a candidate parameter \(\boldsymbol{\theta}^*\) from a proposal distribution \(q(\cdot \mid \boldsymbol{\theta}_t)\), followed by simulating synthetic data \(\mathbf{x}^*\) from the model \(P_{\boldsymbol{\theta}^*}\). The candidate is then accepted with the Metropolis–Hastings acceptance probability  
\[
\alpha\big[(\boldsymbol{\theta}, \mathbf{x}), (\boldsymbol{\theta}^*, \mathbf{x}^*)\big] = \min \left\{1, \frac{\pi(\boldsymbol{\theta}^*) K_\varepsilon(\mathbf{x}^*, \mathbf{y}) q(\boldsymbol{\theta}^* \mid \boldsymbol{\theta})}{\pi(\boldsymbol{\theta}) K_\varepsilon(\mathbf{x}, \mathbf{y}) q(\boldsymbol{\theta} \mid \boldsymbol{\theta}^*)} \right\},
\]  
otherwise the chain remains at the current state. Typically, the proposal distribution \(q(\cdot \mid \boldsymbol{\theta}_t)\) is taken to be a Gaussian distribution centered at the current state \(\boldsymbol{\theta}_t\). Conversely, if the proposal distribution is independent of the current state \(\boldsymbol{\theta}_t\), the resulting move corresponds to a global ABC-MCMC update.

Formally, the transition kernel of the Global-Local ABC-MCMC is given by  
\[
k\big((\boldsymbol{\theta}, \xbold), (\boldsymbol{\theta}^*, \xbold^*)\big) = \gamma \, k_g\big((\boldsymbol{\theta}, \xbold), (\boldsymbol{\theta}^*, \xbold^*)\big) + (1 - \gamma) \, k_l\big((\boldsymbol{\theta}, \xbold), (\boldsymbol{\theta}^*, \xbold^*)\big),
\]  
where \(k_l\) denotes a local ABC-MCMC kernel, and \(k_g\) is a global kernel that proposes more distant candidates to facilitate large jumps across the parameter space.
The parameter \(\gamma \in [0,1]\) controls the balance between exploitation (local moves) and exploration (global moves). When \(\gamma = 0\), the algorithm reduces to the standard ABC-MCMC, relying exclusively on local proposals; when \(\gamma = 1\), only global proposals are used. By integrating these two complementary mechanisms, the Global-Local ABC-MCMC achieves improved mixing and robustness, enabling efficient sampling from challenging, potentially multimodal approximate posteriors. Algorithm \ref{sec:basic} illustrates one iteration of the Global-Local ABC-MCMC. 

\begin{algorithm}[H]
\caption{Single stage of Global-Local ABC-MCMC}
\label{alg:GL-ABC-MCMC}
{\bf Input:} Previous state $(\thetabold_t,\xbold_{t})$, global ABC-MCMC move $k_g\big((\boldsymbol{\theta}_t, \xbold_t), \cdot\big)$, local ABC-MCMC move $k_l\big((\boldsymbol{\theta}_t, \xbold_t), \cdot\big)$.

\begin{algorithmic}[1]
\STATE Sample $u\sim \mathcal{U}[0,1]$.
\IF{$u<\gamma$}
\STATE Obtain $(\thetabold_t,\xbold_{t})$ by running a global ABC-MCMC move $k_g\big((\boldsymbol{\theta}_t, \xbold_t), \cdot\big)$.
\ELSE
\STATE Obtain $(\thetabold_t,\xbold_{t})$ by running a local ABC-MCMC move $k_l\big((\boldsymbol{\theta}_t, \xbold_t), \cdot\big)$.
\ENDIF
\end{algorithmic}
\end{algorithm}

\subsection{The global move based on iterated-sampling importance resample}
\label{subsec:i-SIR}
Directly using a proposal distribution \( q(\cdot) \) that is independent of the current state to construct global ABC-MCMC moves can indeed facilitate transitions between modes. However, when such an independent proposal diverges significantly from the ABC posterior, the Metropolis–Hastings acceptance probability $\alpha\left[(\thetabold,\xbold),(\thetabold^*,\xbold^*)\right]=\operatorname{min}\left\{1, \frac{\pi(\thetabold^*)K_\varepsilon(\xbold^*,\ybold)q(\thetabold)}{\pi(\thetabold)K_\varepsilon(\xbold,\ybold)q(\thetabold^*)}\right\}$ tends to be very low. This limitation poses challenges for the global exploration capabilities of Global-Local ABC-MCMC, especially in high-dimensional parameter spaces. Therefore, it is essential that the global proposal closely approximates the ABC posterior; however, inferring such a proposal is nearly as complex as inferring the ABC posterior itself. 

To address this challenge, we propose an effective global ABC-MCMC move constructed via ABC \textit{Sampling Importance Resampling} (SIR). SIR is a commonly used technique to transition from a simple distribution to a more complex target distribution. Specifically, the ABC-SIR procedure involves drawing samples \(\thetabold_i\) from a simple and easy-to-sample proposal distribution \(q(\thetabold)\), and for each \(\thetabold_i\), generating corresponding simulated data \(\xbold_i \sim P_{\thetabold_i}\), where \(i=1,\ldots,N\). Each of these \(N\) samples is then assigned an importance weight
\[
W_i \propto \frac{\pi(\thetabold_i) K_\varepsilon(\xbold_i, \ybold)}{q(\thetabold_i)}.
\] 
Finally, resampling from these weighted samples yields a set of samples approximately drawn from the target ABC posterior distribution \(\pi_\varepsilon(\thetabold \mid \ybold)\).

Here we utilize \textit{iterated} SIR (i-SIR) to design efficient global ABC-MCMC move. The likelihood-based i-SIR is originally proposed in \cite{tjelmeland2004using} and further studied in \cite{andrieu2010particle, andrieu2018uniform, samsonov2022local-global}. Here, we introduce a likelihood-free version of i-SIR as a more efficient global ABC-MCMC moves, refer to as ABC-i-SIR. Algorithm \ref{alg:i-SIR} illustrates a single stage of ABC-i-SIR. In one stage of ABC-i-SIR,  we sample $\left\{\thetabold_{(j)}^*\right\}_{j=1:N_{b}}$ from the independent proposal distribution, and generate corresponding synthetic data $\left\{\xbold_{(j)}^*\right\}_{j=1:N_{b}}$. We then combine the newly generated particles with the last state of the chain $(\thetabold_{(0)}^*,\xbold_{(0)}^*) = (\thetabold_{t},\xbold_{t})$ as candidate particles. The new state $(\thetabold_{t+1},\xbold_{t+1})$ is obtained by sampling from the candidate particles $\left\{\thetabold_{(j)}^*, \xbold_{(j)}^*\right\}_{j=0:N_{b}}$ with importance weights $w_j\propto\pi(\thetabold_{(j)}^*)K_\varepsilon(\xbold_{(j)}^*,\ybold)/q(\thetabold_{(j)}^*)$.
\begin{algorithm}[H]
\caption{Single stage of ABC-i-SIR algorithm with independent proposals}
\label{alg:i-SIR}
{\bf Procedure} ABC-i-SIR($(\thetabold_t,\xbold_{t})$, $q$, $\pi$, $P_\thetabold$, $K_\varepsilon$)

{\bf Input:} Previous state $(\thetabold_t,\xbold_{t})$, proposal $q(\cdot)$, simulator $P_\thetabold$, kernel function $K_\varepsilon$, prior $\pi(\thetabold)$.

{\bf Output:} New state $(\thetabold_{t+1}$,$\xbold_{{t+1}})$.
\begin{algorithmic}[1]
\STATE Set $(\thetabold_{(0)}^*,x_{(0)}^*)=(\thetabold_t,\xbold_{t})$, draw $\thetabold_{(1:N_{b})}^*$ from $q(\cdot)$.
\STATE Generate synthetic data $\xbold_{(i)}^*$ from $P_{\thetabold_{(i)}^*}$ for $i=1,\ldots, N_{b}$.
\STATE Compute weight $w_{i}\propto\pi(\thetabold_{(i)}^*)K_\varepsilon(\xbold_{(i)}^*,\ybold)/q({\thetabold_{(i)}^*})$, for $i=0,\ldots,N_{b}$.
\STATE Sample $(\thetabold_{t+1},\xbold_{{t+1}})$ from $\left\{(\thetabold_{(i)}^*,\xbold_{(i)}^*)\right\}_{i=0:N_{b}}$ with weight $\left\{w_{i}\right\}_{i=0:N_{b}}$.
\end{algorithmic}
\end{algorithm}

A simple choice of the importance proposal in ABC-i-SIR is the prior distribution. For complex models, to improve algorithmic efficiency, we can select a larger threshold $\varepsilon^* > \varepsilon$ and perform a small number of simulations to obtain a rough estimate of the posterior $\pi_{\varepsilon^*}$, which is then used as the importance proposal for ABC-i-SIR.

In Proposition \ref{prop: two-stage Gibbs} of Section \ref{sec:prop}, we show that the ABC-i-SIR algorithm can be interpreted as a systematic-scan two-stage Gibbs sampler. Given that candidate parameters are drawn from a proposal distribution independent of the current state, each iteration of ABC-i-SIR effectively performs a global ABC-MCMC update.  Moreover, Proposition \ref{prop: iSIR Uniform geometric ergodicity} in Section \ref{sec:prop} establishes the uniform geometric ergodicity of the MCMC kernel induced by ABC-i-SIR.

\subsection{A gradient based local proposal}
In practice, it is common to employ a $p$-dimensional normal distribution as proposal for local MCMC steps (\eg~$q(\thetabold^*\mid\thetabold)=\mathcal{N}(\thetabold^*;\thetabold,\sigma^2 \boldsymbol{I}_p)$).
The scale parameter of the normal distribution significantly influences the efficiency and performance of the algorithm, particularly for high-dimensional parameter space. Designing an appropriate scale poses a challenge. 

It's necessary to design a general proposal mechanism which provides large proposal transitions with high probability of acceptance. \textit{Metropolis adjusted Langevin algorithm} (MALA) \citep{Grenander1994MALA,roberts1996exponential,roberts1998optimal,girolami2011riemann} combines Langevin dynamics with a Metropolis-Hastings correction to create an adaptive and efficient MCMC algorithm. 

 The Langevin diffusion process is defined by the stochastic differential equation (SDE)
\begin{equation}
    d\thetabold(t)=\nabla_\thetabold \mathcal{L}\{\thetabold(t)\}dt/2+d\boldsymbol{W}(t),
\end{equation}
where $\mathcal{L}(\thetabold)$ denotes the logarithm of target distribution, $\boldsymbol{W}$ denotes a $p$-dimensional Brownian motion. The diffusion is irreducible, strong Feller and aperiodic, with stationary distribution $\pi_T(\cdot)\propto\exp(\mathcal{L}(\cdot))$. This process can be regarded as a continuous-time sampling method. Unfortunately, the implementation is infeasible in practice. The common method is to use 
discretized approximations via Euler-Maruyama discretization:
\begin{equation}
    \thetabold^*=\thetabold + \eta^2\nabla_\thetabold\mathcal{L}(\thetabold)/2+\eta z,
\end{equation}
where $\eta$ is the step-size and $z\sim \mathcal{N}(0,\boldsymbol{I}_p)$.  However, when discretizing the diffusion, some bias is introduced, then convergence to the invariant distribution $\pi_T$ is no longer guaranteed. MALA employs a Metropolis-Hastings rejection-acceptance step after every iteration to ensure the convergence property. The discrete form of Langevin diffusion plays the role of an instrumental proposal distribution. In particular,  $q\left(\boldsymbol{\theta}^* \mid \boldsymbol{\theta}\right)=\mathcal{N}\left\{\boldsymbol{\theta} \mid \thetabold+\eta^2\nabla_\thetabold\mathcal{L}(\thetabold)/2, \eta^2 \boldsymbol{I}_p\right\}$ and the corresponding acceptance probability is $\alpha(\thetabold,\thetabold^*)=\min\left\{1,\frac{\exp(\mathcal{L}(\thetabold^*))p(\thetabold\mid\thetabold^*)}{\exp(\mathcal{L}(\thetabold))p(\thetabold^*\mid\thetabold)}\right\}$.

Recall that our target distribution is $\pi_\varepsilon(\thetabold\mid\ybold)\propto\pi(\thetabold)p_\varepsilon(\ybold\mid\thetabold)$, where $p_\varepsilon(\ybold\mid\thetabold)=\int p(\xbold\mid\thetabold) K_\varepsilon(\xbold,\ybold)d\xbold$ is the approximate likelihood. Then $\mathcal{L}(\thetabold)=\log(\pi(\thetabold))+\log(p_\varepsilon(\ybold\mid\thetabold))$. 
The core of ABC-MALA is computing the gradient of log-likelihood. \cite{meeds2015HABC} apply HMC to ABC and propose to compute the gradient using \textit{common random numbers} (CRN). The key idea of CRN is to represent $\xbold$ as a deterministic function of variables $\thetabold$ and $\omega$, that is to define the random variable $\xbold$ from $p(\xbold\mid\thetabold)$ as a deterministic function $f(\thetabold,\omega)$, where $\omega$ is a random seed. Then, the estimated likelihood based on $\varepsilon$-kernel is
\begin{equation}
p_\varepsilon(\ybold\mid\thetabold)=\sum_{s=1}^SK_\varepsilon(f({\thetabold},\omega_{s}),\ybold),
\end{equation}
where $\omega_s$ ($s=1,\ldots,S$) are $S$ random seeds.
\cite{meeds2015HABC} suggest to approximate the conditional distribution $p_\varepsilon(\ybold\mid\thetabold)$ by $K_\varepsilon(f(\thetabold,\omega_m),\ybold)$ under the assumption that the threshold is very close to $0$. Here, $\omega_m$ represents a seed that ensures $K_\varepsilon(f(\thetabold,\omega_m),\ybold)$ is the largest among $\left\{K_\varepsilon(f(\thetabold,\omega_i),\ybold)\right\}_{i=1:S}$. Then, the gradient of log-likelihood is estimated by
\begin{equation}
    \label{eq:CRNmax}\nabla_\thetabold\log(p_\varepsilon(\ybold\mid\thetabold))=\frac{1}{2 d_\thetabold}\left\{\log(K_\varepsilon(f({\thetabold}+ d_\thetabold,\omega_{m}^+),\ybold))-\log(K_\varepsilon(f({\thetabold}-d_ \thetabold,\omega_{m}^-),\ybold))\right\},
\end{equation}
where $\omega_m^+=\arg\max_{\omega_{i}\in\{\omega_{1:S}\}}K_\varepsilon(f({\thetabold}+ d_\thetabold,\omega_{i}),\ybold)$ and $\omega_m^-=\arg\max_{\omega_{i}\in\{\omega_{1:S}\}}K_\varepsilon(f({\thetabold}- d_\thetabold,\omega_{i}),\ybold)$. In this article, we denote the method as $\text{CRN}_{max}$. In practice, ensuring the efficiency of ABC methods with a very small $\varepsilon$ is challenging. Here, we estimate the partial derivative with respect to $\thetabold$ without the small threshold assumption as follows:
\begin{equation}
\nabla_\thetabold\log(p_\varepsilon(\ybold\mid\thetabold))=\frac{1}{2 d_\thetabold}\left(\log\left(\sum_{s=1}^SK_\varepsilon(f({\thetabold}+ d_\thetabold,\omega_{s}),\ybold)\right)-\log\left(\sum_{s=1}^SK_\varepsilon(f({\thetabold}-d_ \thetabold,\omega_{s}),\ybold)\right)\right),
\label{eq:CRN_mean}
\end{equation}
where $d_\thetabold$ is a small perturbation.
The detailed finite difference stochastic approximation with CRN ($\text{CRN}_{mean}$) is shown in the Appendix A.1.

However, in the tail region of the posterior distribution, the logarithmic function can exacerbate the errors associated with Monte Carlo integration, leading to significantly inaccurate gradient estimates. In this study, we employ a Gaussian distribution to model the likelihood based on simulated data to obtain a more stable and accurate estimator when there are few simulations to estimate the integral. By utilizing the Gaussian kernel as the kernel density function, we derive the gradient estimate as follows:
\begin{equation}
\label{eq:grandient-gaussian}
\nabla_\thetabold\log(p_\varepsilon(\ybold\mid\thetabold))=\frac{1}{2d_\thetabold}\left(-\frac{1}{2}\log\left(\frac{\hat{\sigma}_{\thetabold^+}^2+\varepsilon^2}{\hat{\sigma}_{\thetabold^-}^2+\varepsilon^2}\right)-\frac{(\ybold-\hat{\mu}_{\thetabold^{+}})^2}{2(\hat{\sigma}_{\thetabold^+}^2+\varepsilon^2)}+\frac{(\ybold-\hat{\mu}_{\thetabold^{-}})^2}{2(\hat{\sigma}_{\thetabold^-}^2+\varepsilon^2)}\right),
\end{equation}
where the mean $\hat{\mu}_{\thetabold^{+}}$ and variance $\hat{\sigma}_{\thetabold^+}^2$  are estimated from samples drawn from the distribution $p(\cdot\mid\thetabold+d\thetabold)$, while mean $\hat{\mu}_{\thetabold^{-}}$ and variance $\hat{\sigma}_{\thetabold^-}^2$ are derived similarly. Samples from $p(\cdot\mid\thetabold+d\thetabold)$ and $p(\cdot\mid\thetabold-d\thetabold)$ are generated using a common random seed.
\section{Two adaptive schemes}
\label{sec:adaptive}

\subsection{ An adaptive proposal of ABC-i-SIR}
In Section \ref{subsec:i-SIR}, we employ the likelihood-free SIR technique iteratively to construct an efficient global kernel for ABC-MCMC.  The proposal distribution plays a crucial role in importance sampling, influencing the efficiency of the sampling process. In essence, the proposal determines where samples are drawn from, and its similarity to the target distribution significantly impacts the quality of estimates. An effective proposal reduces variability and yields more reliable results. The calculation of weights depends on how well the proposal aligns with the target, and optimal performance is achieved when the proposal closely approximates the target distribution.  In this section, we propose an adaptive procedure that iteratively enhances the importance proposal distribution within the ABC-i-SIR algorithm, aiming to progressively align it more closely with the target posterior distribution.

In this work, we leverage the idea of sequential importance sampling combined with a flexible density estimation framework based on weighted samples to progressively improve the proposal distribution. Specifically, we introduce a decreasing sequence of tolerance thresholds 
\[
\varepsilon^*_1 \geq \varepsilon^*_2 \geq \cdots \geq \varepsilon^*_T,
\]
and define a series of intermediate posterior distributions 
\[
\pi_{\varepsilon^*_1}(\thetabold \mid \ybold), \; \pi_{\varepsilon^*_2}(\thetabold \mid \ybold), \; \ldots, \; \pi_{\varepsilon^*_T}(\thetabold \mid \ybold).
\]
Rather than directly approximating the target ABC posterior, 
we gradually update the proposal distribution to approach the target posterior through these intermediate distributions. The discrepancy between the proposal and the approximate target at each stage is reduced. This leads to a larger effective sample size of weighted samples and consequently leads to improved approximation quality. This series of intermediate target distribution is often used in ABC sequential Monte Carlo methods \citep{del2012adaptive}.

Concretely, the \(j\)-th update of the importance proposal distribution is constructed using the weighted sample set \(\left\{\left(\thetabold_{(i)}^*, w_{\varepsilon_j^*,i}^*\right)\right\}_{i=1}^{n_t}\), where the normalized importance weights are given by  
\begin{equation}
    w_{\varepsilon_j^*,i}^* \propto \frac{\pi(\thetabold_{(i)}^*) \, K_{\varepsilon_j^*}(\xbold_{(i)}^*, \ybold)}{q_{\varepsilon_{j-1}^*}^*(\thetabold_{(i)}^*)}.
    \label{eq: weight-prop-update}
\end{equation}
Here \(\{(\thetabold_{(i)}^*, \xbold_{(i)}^*)\}_{i=1}^{n_t}\) denotes a  set of \(n_t\) training data pairs sampled from the ABC-i-SIR algorithm, where the proposal distribution at the previous iterations is \(q_{\varepsilon_{j-1}^*}\). The numerator, \(\pi(\thetabold_{(i)}^*) \, K_{\varepsilon_j^*}(\xbold_{(i)}^*, \ybold)\), represents the unnormalized intermediate target distribution \(\pi_{\varepsilon_{j}^*}((\thetabold, \xbold) \mid \ybold)\).

To avoid increasing the simulation cost while ensuring a sufficient number of samples, we run a single Markov chain and update the importance proposal distribution once every \(S\) ABC-i-SIR iterations. Specifically, \(n_t = S \times N_b\) weighted samples are collected for each update, where \(N_b\) denotes the batch size of each ABC-i-SIR iteration.

Kernel density estimation (KDE) \citep{terrell1992variable} serves as a natural and computationally efficient choice for this density estimation step:  
\[
q_{\varepsilon_j^*}^*(\thetabold) = \sum_{i=1}^{n_t} w_{\varepsilon_j^*,i}^* \, \tilde{K}_h(\thetabold, \thetabold_{(i)}^*),
\]  
where \(\tilde{K}_h\) represents a smooth kernel with bandwidth \(h\). Nonetheless, KDE is not the sole viable estimator; alternative approaches such as Gaussian mixture models \citep{mclachlan2014number}, adaptive kernel methods \citep{van2003adaptive}, or generative models grounded in machine learning frameworks \citep{prop-goodfellow2014generative,prop-kingma2013auto,papamakarios2021normalizing} can be employed to capture complex posterior geometries with potentially greater flexibility.

Moreover, to ensure the safety and robustness of the proposal distribution, an auxiliary tolerance threshold \(\varepsilon^*_T\) can be set larger than the final target threshold \(\varepsilon\).  To guarantee the proposal distribution steadily and reliably approaches the target posterior, we adopt an adaptive scheme that gradually reduces the auxiliary tolerance while maintaining a sufficient number of nonzero-weighted training samples. Specifically, the sequence \(\{\varepsilon^*_j\}\) is obtained by solving
\begin{equation}
PA\left(\left\{\xbold_{(i)}^* \right\}_{i=1:n_t},\varepsilon^*_j\right)
    =\alpha \, PA\left(\left\{\xbold_{(i)}^* \right\}_{i=1:n_t},\varepsilon^*_{j-1}\right),
    \label{eq:eps-update}
\end{equation}
where $$PA\left(\left\{\xbold_{(i)}^* \right\}_{i=1:n_t},\varepsilon^*_j\right)=
\sum_{i=1}^{n_t}\mathbb{I}\left(\mathfrak{D}(\xbold_{(i)}^*,\ybold)<\varepsilon^*_j\right)$$ denotes the number of alive particles under tolerance threshold $\varepsilon^*_j$. Here, \(\mathbb{I}(\cdot)\) is the indicator function, \(\mathfrak{D}(\xbold,\ybold)\) measures the discrepancy between the simulated data \(\xbold\) and the observed data \(\ybold\), and \(0 < \alpha < 1\) controls the reduction rate of the proportion of alive samples.
In general, the initial proposal distribution is set to the prior distribution, i.e., \(q_{\varepsilon_0} = \pi\), with the initial tolerance threshold \(\varepsilon_0 = \infty\). In practice, it is recommended to set the parameter \(\alpha\) within the range of 0.6 to 0.9. This range ensures a gradual reduction of the auxiliary threshold while maintaining a sufficiently large number of active particles to effectively update the proposal distribution. Algorithm \ref{alg:Ai-SIR} presents the detailed procedure for one iteration of the adaptive ABC-i-SIR (ABC-Ai-SIR) update.

\begin{algorithm}[H]
\caption{One iteration of the ABC-Ai-SIR}
\label{alg:Ai-SIR}
{\bf Procedure} ABC-Ai-SIR($(\thetabold_t,\xbold_{t}),P_\thetabold,K_\varepsilon,\varepsilon^*_j,q_{\varepsilon^*_j},\varepsilon^*_T,\alpha,\mathcal{D}$).

{\bf Input:} Previous state $(\thetabold_t,\xbold_{t})$, simulator $P_\thetabold$, kernel $K_\varepsilon$, prior $\pi(\thetabold)$, number of steps to collect data $S$, auxiliary threshold $\varepsilon^*_j$, importance proposal density $q_{\varepsilon^*_j}$, target auxiliary threshold $\varepsilon^*_T$, auxiliary threshold reduction factor $\alpha$, currently collected training data $\mathcal{D}$.

{\bf Output:} New state $(\thetabold_{t+1},\xbold_{{t+1}})$.

\begin{algorithmic}[1]
\STATE Set $\thetabold_{(0)}^*=\thetabold_t$, and $\tilde{w}_0=\pi(\thetabold_{t})K_{\varepsilon}(\xbold_t,\ybold)/q_j(\thetabold_t)$.
\FOR{$i=1:N_b$}
\STATE Draw $\thetabold_{(1:N_{b})}^*$ from $p_t(\cdot)$ and generate simulate data $\xbold_{(i)}^*$ from $P_{\thetabold_{(i)}^*}$.
\STATE Compute weight $\tilde{w}_i=\pi(\thetabold_{(i)}^*)K_{\varepsilon}(\xbold_{(i)}^*,\ybold)/q_j(\thetabold_{{(i)}}^*)$. 
\ENDFOR
\STATE Sample $\thetabold_{t+1}$ from $\left\{\thetabold_{(i)}^*\right\}_{i=0:N_{b}}$ with weight $\left\{w_{i}\right\}_{i=0:N_{b}}$, where $w_i=\frac{\tilde{w}_i}{\sum_{i=1}^{N_b}\tilde{w}_i}$ for $i=1,\cdots,N_b$.
\STATE Collect $\left\{(\thetabold_{(i)}^*,\xbold_{(i)}^*)\right\}_{i=1:N_b}$ into the data set $\mathcal{D}$.
\IF{$|D|=SN_b$}
\STATE $j=j+1$.
\STATE If $\varepsilon^*_{j-1}>\varepsilon^*_T$, obtain $\varepsilon^*_{j}$ by solving Eq. \eqref{eq:eps-update}, then set $\varepsilon^*_{j}=\max\{\varepsilon^*_T,\varepsilon^*_j \}$.
\STATE Obtain $q_{\varepsilon^*_{j}}$ by update $q_{\varepsilon^*_{j-1}}$ by using weighted data $\left\{\left(\thetabold_{(i)}^*,w_{\varepsilon^*_{j},i}^*\right)\right\}_{i=1:SN_b}$ in the data set $\mathcal{D}$, where weight $w_{\varepsilon^*_{j},i}^*$ is defined in Eq. \eqref{eq: weight-prop-update}
\STATE Initialize $\mathcal{D}$ to an empty set.
\ENDIF
\end{algorithmic}
\end{algorithm}

\subsection{Sequential optimization of hyper-parameters}
In order to ensure the convergence rate of GL-ABC-MCMC, we need to optimize the hyperparameters (\ie~the global frequency $\gamma$, the batch size $N_b$). The expected squared jump distance (ESJD) \citep{pasarica2010adaptivelyESJD,atchade2011towardsESJD,roberts2014minimisingESJD,yang2020optimalESJD} is utilized to adjust the proposal distribution in the MCMC algorithm to improve the algorithm's mixing and exploration. 
The ESJD in one dimension is defined as
$$\text{ESJD}=E[|\thetabold_{t+1}-\thetabold_{t}|]=2(1-\rho_1)\operatorname{Var}_{\pi_\varepsilon}(\thetabold),$$
where $\rho_1$ is the first-order auto-correlation of the Markov chain. Maximizing
ESJD is equivalent to minimizing the first-order auto-correlation and thus maximizing the efficiency if the higher order auto-correlations are monotonically increasing with respect to the first-order auto-correlation \citep{pasarica2010adaptivelyESJD}. \cite{roberts2014minimisingESJD} demonstrated that maximizing the ESJD is equivalent to minimizing the asymptotic variance of diffusion limits of MCMC methods in some certain scenarios.

\cite{yang2020optimalESJD} consider maximizing ESJD to optimal scaling of random-walk Metropolis algorithms on general target distributions, and show the asymptotically optimal acceptance rate $0.234$ can be obtained
under general realistic sufficient conditions on the target distribution. 

For high-dimensional problems, \cite{pasarica2010adaptivelyESJD} defined ESJD as \begin{equation}
\label{eq: ESJD_sigma}
    \text{ESJD} = E[\|\thetabold_{t+1} - \thetabold_t)\|_{\Sigma^{-1}}^2]= 2\operatorname{tr}\left(I - \Sigma^{-1}E\left[(\thetabold_{t+1} - \bar{\thetabold})(\thetabold_{t} - \bar{\thetabold})^\top\right]\right),
\end{equation}
where $\bar{\thetabold}$ and $\Sigma$ are the mean and covariance matrix of $ \thetabold $ from the stationary distribution. Maximizing \eqref{eq: ESJD_sigma} is equivalent to minimizing the first-order auto-correlation of the Markov chain. 
However, the covariance matrix $\Sigma$ is typically unknown until the target distribution is well-estimated. 
In this paper, we define ESJD based on the $D$-criterion as follows
\begin{equation}
\label{eq: ESJD_D}
    \text{ESJD} = \left| \operatorname{E} \left[(\thetabold_{t+1} - \thetabold_t)(\thetabold_{t+1} - \thetabold_t)^\top\right]\right|^{\frac{1}{p}} = \left(|2\Sigma|\left|I-\Sigma^{-1}E\left[(\thetabold_{t+1} - \bar{\thetabold})(\thetabold_{t} - \bar{\thetabold})^\top\right]\right|\right)^{\frac{1}{p}},
\end{equation}
where $p$ is the dimension of $\thetabold$.
This accounts for different scales of the parameter dimensions while does not require the estimation of $\Sigma$. In extreme cases, Equation \eqref{eq: ESJD_D} takes the maximum value $|2\Sigma|^{1/p}$ when $\thetabold_{t+1}$ and $\thetabold_{t}$ are independent, and takes the minimum value 0 when $\thetabold_{t+1}$ and $\thetabold_{t}$ are equal. When $\Sigma$ is a diagonal matrix, Equation \eqref{eq: ESJD_D} measures the geometric mean of one dimension ESJD from different dimensions. 
Furthermore, considering the cost associated with different proposal mechanisms, we define a unit cost version of ESJD (cESJD) that takes the computational cost into account as: 
\begin{equation}
    \text{cESJD} = \frac{\text{ESJD}}{\mathcal{C}},
\end{equation}
where $ \mathcal{C} $ denotes the average cost per MCMC iteration (\eg~computing time).
This ensures a balanced selection of hyper-parameters (\ie~$\gamma$ and $N_b$)  that not only maximizes the exploration of the parameter space but also accounts for the efficiency of the MCMC algorithm in terms of computational resources.

To enhance the performance of the GL-ABC-MCMC algorithm, we propose using a sequential optimization algorithm to select an optimal hyper-parameter combination. This process involves selecting a set of candidate points in the hyper-parameter space based on specific rules, such as informative priors or non-informative uniform density. For each hyper-parameter combination, a short GL-ABC-MCMC algorithm is run, and the ESJD value is calculated. The best hyper-parameter combination, which yields the highest ESJD, is then used as the center of the parameter space, and the parameter space is narrowed down accordingly. This process is repeated until convergence is achieved or a predetermined requirement is met. 

When there is a lack of prior information about hyper-parameters, using uniform design (UD) \citep{fang1993number, fang2018UDbook} to explore the hyper-parameter space is robust and efficient. UD is constructed based on the number-theoretic method or quasi-Monte Carlo method, and it possesses good space-filling properties. 
\cite{yang2021hyperparameter} proposed a method for hyper-parameter optimization via sequential uniform designs. By virtue of its space-filling property, uniform design ensures adequate exploration of the entire parameter space. Compared to random sampling, uniform design can achieve similar exploration effects with fewer sample points, thereby saving experimental resources and costs.

\begin{algorithm}[H]
\caption{Sequential optimization algorithm}
\label{alg:sequential}

{\bf Input:} Hyper-parameter space $\mathcal{K}$, maximum number of replications $R$, number of hyper-parameter combinations per stage $n_s$, iteration number of pilot GL-ABC-MCMC run $N_p$.

{\bf Output:} The optimal hyper-parameter combination  $\kappa^*$.
\begin{algorithmic}[1] 
\STATE Initialize hyper-parameter set $ \mathbb{K}=\emptyset$ and $\operatorname{ESJD}^*=0$.
\FOR{$r=1,\ldots, R$}
\STATE Select $\kappa_{1:n_s}$ in hyper-parameter space, $\kappa_i\notin \mathbb{K}$ for $i=1,\ldots,n_s$, and update $\mathbb{K}=\mathbb{K}\bigcup\kappa_{1:n_s}$.
\STATE For each $\kappa_i$, run GL-MCMC with $N_p$ iterations and compute $\operatorname{ESJD}_{1:n_s}$.
\STATE If $\operatorname{ESJD}_{m}> \operatorname{ESJD}^*$, update $\kappa^*= \kappa_m$ and $\operatorname{ESJD}^*=\operatorname{ESJD}_{m}$, where $m=\arg\max_i \{\operatorname{ESJD}_{1:n_s}\}$.
\ENDFOR
\end{algorithmic}
\end{algorithm}
The detailed algorithm is shown in Algorithm \ref{alg:sequential}. The algorithm achieves adaptive exploration of the hyper-parameter space by sequentially adjusting the search space. At each stage, the search space is automatically adjusted to focus on the most promising region.

\section{Properties}
\label{sec:prop}

In this section, we demonstrate that the MCMC kernels of both ABC-i-SIR and GL-ABC-MCMC are $V$-uniformly geometrically ergodic under certain mild conditions.

\noindent\textbf{Notations:} Denote ${\mathbb{N}^*}=\mathbb{N} \backslash\{0\}$. For a measurable function $f: \mathbb{X} \mapsto \mathbb{R}$, we define $|f|_{\infty}=$ $\sup _{x \in \mathbb{X}}|f(x)|$ and $\pi(f):=\int_{\mathbb{X}} f(x) \pi({d} x)$. For a function $V: \mathbb{X} \mapsto[1, \infty)$, we introduce the $V$-norm of two probability measures $\xi$ and $\xi^{\prime}$ on $(\mathbb{X}, \mathcal{X}),\left\|\xi-\xi^{\prime}\right\|_V:=\sup _{|f(x)| \leq V(x)}\left|\xi(f)-\xi^{\prime}(f)\right|$. If $V \equiv 1,\|\cdot\|_1$ is equal to the total variation distance (denoted $\|\cdot\|_{{TV}}$).  For simplicity, we denote the parameter-data pair   $(\thetabold,\xbold)$ as $\zbold$, $\zbold\in\mathbb{Z}=\Theta\times\mathbb{Y}$, and the  ABC posterior distribution $\pi_\varepsilon(\thetabold,\xbold\mid\ybold)$ as $\pi_\varepsilon(\zbold\mid\ybold)$, which is defined on the measurable space $(\mathbb{Z},\mathcal{Z})$.

\begin{definition}
    A Markov kernel $Q$ with invariant probability measure $\pi$ is $V$-geometrically ergodic if there exist constants $\rho\in(0,1)$ and $M<\infty$ such that, for all $x\in\mathbb{X}$ and $k\in\mathbb{N}$, $\|Q^k(x,\cdot)-\pi\|_V\le M\{V(x)+\pi(V)\}\rho^k$.
\end{definition}

In the ABC-i-SIR algorithm, we extend the proposal distribution from $q(\thetabold)$ to $\lambda(\zbold) \triangleq q(\thetabold)p(\xbold\mid\thetabold)$ by introducing the auxiliary variable $\xbold$, which represents simulated data generated under the parameter $\thetabold$.
The Markov chain generated by ABC-i-SIR admits the following Markov kernel
$${P}_{N_{b}}(\zbold, A) =\int \delta_\zbold\left({d} \zbold_{(0)}\right) \sum_{i=0}^{N_{b}} \frac{w\left(\zbold_{(i)}\right)}{\sum_{j=0}^{N_{b}} w\left(\zbold_{(j)}\right)} 1_{A}\left(\zbold_{(i)}\right) \prod_{j=0}^{N_{b}} \lambda\left({d} \zbold_{(j)}\right),$$
where $w(\zbold)=\pi(\thetabold)p(\xbold\mid\thetabold)K_\varepsilon(\xbold,\ybold)/\lambda(\zbold)$ is the unnormalized importance weight function. We denote the function as $w(\zbold)=\pi_\varepsilon(\zbold\mid\ybold)\cdot\lambda(w)/\lambda(\zbold)$, where $\lambda(w)$ is the normalizing constant of the distribution  $\pi_\varepsilon(\zbold\mid\ybold)$. Following the work of \cite{samsonov2022local-global}, we demonstrate that the likelihood-free ABC-i-SIR algorithm can also be interpreted as a systematic-scan two-stage Gibbs sampler, which possesses uniform geometric ergodicity. The formal statements of these results are provided in Proposition~\ref{prop: two-stage Gibbs} and Proposition~\ref{prop: iSIR Uniform geometric ergodicity}, and the proofs are shown in the Appendix.

\begin{proposition}
\label{prop: two-stage Gibbs}
    The ABC-i-SIR algorithm can be interpreted as a systematic-scan two-stage Gibbs sampler, that is $\left\{(\thetabold_{(i)}^*,\xbold_{(i)}^*)\right\}_{i=0:{N_b}}\mid(\thetabold_t,\xbold_{t})\rightarrow(\thetabold_{t+1},\xbold_{{t+1}})\mid\left\{(\thetabold_{(i)}^*,\xbold_{(i)}^*)\right\}_{i=0:{N_b}}$, and the marginal distribution of $\left((\thetabold,\xbold),\left\{(\thetabold_{(i)}^*,\xbold_{(i)}^*)\right\}_{i=0:{N_b}}\right)$ with respect to $\left(\thetabold,\xbold\right)$ is $\pi_\varepsilon(\thetabold,\xbold\mid\ybold)$.
\end{proposition}

\begin{proposition}
\label{prop: iSIR Uniform geometric ergodicity}
    If $|w|_\infty<\infty$, for any initial distribution $\xi$ on $(\mathbb{Z},\mathcal{Z})$ and $k\in\mathbb{N}$,
    $\|\xi P_{N_b}^k-\pi_\varepsilon\|_{TV}\le \rho_{N_b}^k$ with $\rho_{N_b} = 1-e_{N_b}$, $e_{N_b}={N_b}/(2L+{N_b}-1)$, and $L=|w|_\infty/\lambda(w)$.
\end{proposition}

In Proposition \ref{prop: iSIR Uniform geometric ergodicity}, the assumption $|w|_\infty<\infty$ implies that the importance proposal must cover the target distribution; that is, $q(\thetabold)>0$ for all $\thetabold\in \{\thetabold:\pi_\varepsilon(\thetabold\mid\ybold)>0\}$, which is a standard assumption for importance sampling. In practice, it suffices to set the importance proposal to cover the prior region. 

It is well established that if both the global and local Markov kernels are geometrically ergodic, then their combination within the Global-Local ABC-MCMC framework inherits geometric ergodicity. In Proposition \ref{prop: iSIR Uniform geometric ergodicity}, we have shown that the Markov kernel underlying the ABC-i-SIR algorithm exhibits uniform geometric ergodicity. 
Building on this, Theorem \ref{Theorem: Mix-convergence} further establishes that when the ABC-i-SIR algorithm is used as the global Markov kernel, even if the local kernel does not satisfy \(V\)-geometric ergodicity, the resulting Global-Local ABC-MCMC kernel still attains \(V\)-geometric ergodicity. In other words, the mixing rate of the Global-Local ABC-MCMC kernel surpasses that of the local kernel alone, indicating improved convergence behavior. 

\begin{assumption}
\label{Ass: local Markov kernel}
    \begin{enumerate}[(i)]
        \item The local Markov kernel $Q$ has $\pi_\varepsilon$ as its unique invariant distribution; 
        \label{Ass: local Markov kernel-1}
        \item There exists a function $V: \zbold \rightarrow [1,\infty)$, such that for all $r \ge r_Q > 1$ there exist $\zeta_{Q,r} \in [0, 1)$, $b_{Q,r} < \infty$, such 
that $QV(\zbold) \le \zeta_{Q,r}V (\zbold) + b_{Q,r}\mathbb{I}_{V_r}$, where $V_r = \{\zbold: V (\zbold) \le r\}$.
\label{Ass: local Markov kernel-2}
    \end{enumerate}  
\end{assumption}
\begin{assumption}
\label{Ass: weight}
        For all $r\ge r_Q$, $w_{\infty,r} := \sup_{\zbold\in V_r}\{w(\zbold)/\lambda(w)\}<\infty$ and $\operatorname{Var}_\lambda(w)/\{\lambda(w)\}^2\le\infty$.
\end{assumption}

\begin{theorem}
\label{Theorem: Mix-convergence}
    Under the Assumptions \ref{Ass: local Markov kernel} and \ref{Ass: weight},  then for all $z\in\mathbb{Z}$, $k\in \mathbb{N}$, and $N_b\ge 2$
    \begin{equation}
        \|K_{N_b}^k(\zbold,\cdot)-\pi_\varepsilon(\cdot)\|_V\le c_{K_{N_b}}\{\pi_\varepsilon(V)+V(\zbold)\}\rho_{K_{N_b}}^k,
    \end{equation}
    with $$
\begin{aligned}
& \log \rho_{K_{N_b}}=\frac{\log \left(1-e_{K_{N_b}}\right) \log \bar{\zeta}_{{K}_{N_b}}}{\log \left(1-e_{K_{N_b}}\right)+\log \bar{\zeta}_{{K}_{N_b}}-\log \bar{b}_{{K}_{N_b}}}, \\
& c_{{K}_{N_b}}=1+\bar{b}_{{K}_{N_b}} /\left[(1-e_{K_{N_b}})(1-\bar{\zeta}_{{K}_{N_b}})\right], \\
& \bar{\zeta}_{{K}_{N_b}}=\zeta_{{K}_{N_b}}+2b_{{K}_{N_b}}/(1+r_{{K}_{N_b}}), \quad \bar{b}_{{K}_{N_b}}=\zeta_{{K}_{N_b}} r_{{K}_{N_b}}+{b}_{{K}_{N_b}} .
\end{aligned}
$$
where $K_{N_b}= \gamma P_{N_b}+(1-\gamma)Q$ with $\gamma\in (0,1]$. 
\end{theorem}

It is important to emphasize that Assumption \ref{Ass: local Markov kernel}\eqref{Ass: local Markov kernel-2} requires the local Markov kernel \(Q\) to satisfy a Foster-Lyapunov drift condition with respect to some function \(V\). This condition is commonly met by classical MCMC kernels such as the Metropolis-Hastings algorithm. However, the Foster-Lyapunov drift condition alone does not guarantee the geometric ergodicity of the local kernel. Additionally, Assumption \ref{Ass: weight} imposes an upper bound on the (normalized) importance weights within the level sets \(V_r\). This is a mild and practical condition: for example, when \(\mathbb{Z}=\mathbb{R}^d\) and \(V\) is norm-like, the sets \(V_r\) are compact, and the importance weights \(w(\cdot)\) remain bounded as long as the importance proposal sufficiently covers the prior support. Moreover, the condition \(\operatorname{Var}_\lambda(w)/\{\lambda(w)\}^2 \leq \infty\) ensures bounded variance of the importance weights, which corresponds to a finite \(\chi^2\)-distance between the proposal and target distributions, playing a critical role in the performance of Sampling Importance Resampling (SIR) methods.

From a practical perspective, the global ABC-i-SIR kernel facilitates strong global moves by enabling efficient jumps across distant regions of the state space, while the local kernel focuses on detailed local exploration. Thus, even if the local kernel alone lacks \(V\)-geometric ergodicity, the combined Global-Local kernel can achieve overall \(V\)-geometric ergodicity, thereby enhancing the efficiency and robustness of the ABC-MCMC algorithm.

\section{Toy examples}
\label{sec:simu}
\subsection{Comparison of different gradient estimation methods}
We use a simple one-dimensional problem to illustrate the effectiveness of the gradient estimation method proposed in this article. We consider a model with the prior density $\pi(\theta)=\mathcal{N}(\theta; 0,1)$, the likelihood function $p(y\mid\theta)=\mathcal{N}(y;\theta,0.01)$, the observation $y_{obs}=0$ and the weighted kernel function $K_\varepsilon(x,y_{obs})=\mathcal{N}(x;y_{obs},\varepsilon^2)$,  where $\mathcal{N}(\theta,\mu,\sigma^2)$ denotes a  Gaussian density with mean $\mu$ and variance $\sigma^2$. Then the closed expression of estimated likelihood and gradient based on $\varepsilon$-kernel are $p_\varepsilon(y\mid\theta)=\mathcal{N}(y,\theta,\varepsilon^2+0.01)$, and $\nabla_\theta\log p_\varepsilon(y\mid\theta)=(y-\theta)/(\varepsilon^2+0.01)$. 

\begin{figure}[H]
    \centering
    \subfigure[$\varepsilon=0.1$]{
    \includegraphics[width=0.4\linewidth]{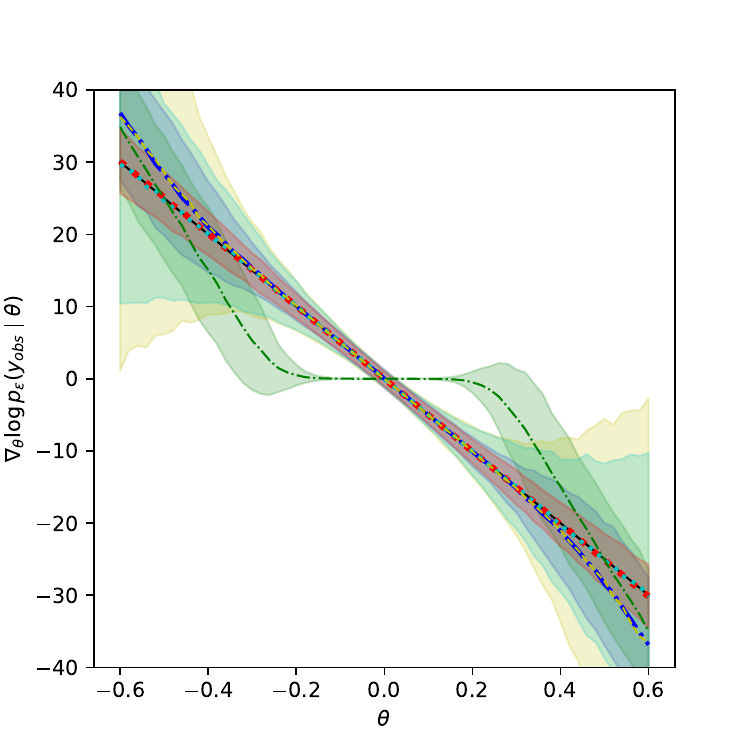}}
    \hspace{0.005\linewidth}
    \subfigure[$\varepsilon=0.05$]{
    \includegraphics[width=0.4\linewidth]{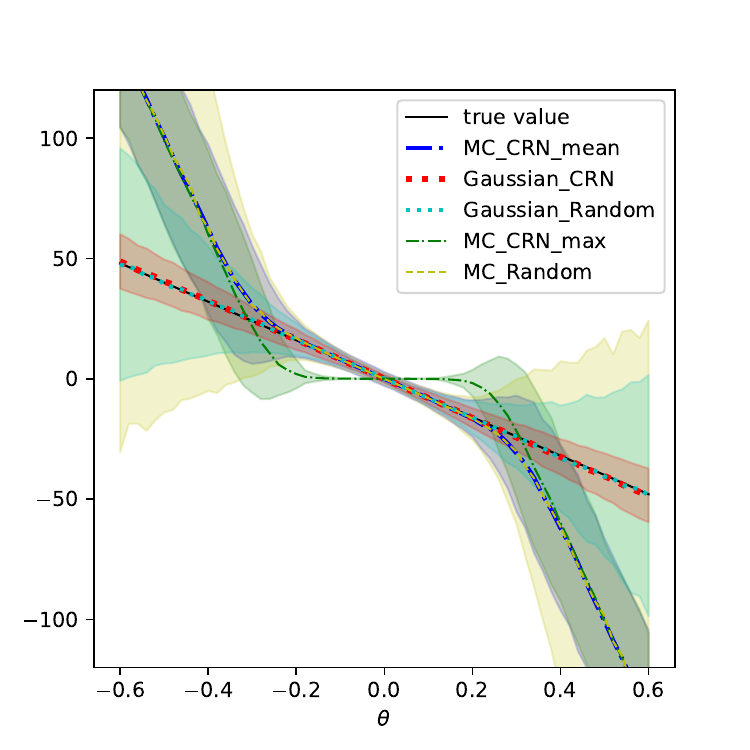}}
    \caption{Comparison of gradient estimation methods. The shaded curved region represents $2\sigma$ of estimated gradient. 
    The standard method (MC\_Random), the method  of \cite{meeds2015HABC} (MC\_CRN\_max, {Eq.}~\eqref{eq:CRNmax}), the improved method based on \cite{meeds2015HABC} (MC\_CRN\_mean, {Eq.}~\eqref{eq:CRN_mean}), the likelihood fitting using Gaussian distribution based on simulation data with CRNs (Gaussian\_CRN, {Eq.}~\eqref{eq:grandient-gaussian} with CRNs), and the likelihood fitting using Gaussian distribution based on random simulation data (Gaussian\_Random, {Eq.}~\eqref{eq:grandient-gaussian}) are compared.}
    \label{fig:grandient}
\end{figure}

 In the Appendix C.1, we demonstrate that for a fixed random seed $\omega_s$, the simulation data $y$ is deterministic for a given $\theta$ and varies smoothly with changes of $\theta$, despite the inherent noise in the simulator. For numerical methods, we fixed $d_{\theta}=0.05$. Figure \ref{fig:grandient} provides gradient estimates obtained from various methods. For each parameter value, gradients are estimated using 100 simulated data points, and repeat the process 1,000 times to calculate both the mean and standard deviation of the gradient estimates. 

Compared to the standard method (random sampling) and the $CRN_{max}$ approach \citep{meeds2015HABC}, the $CRN_{mean}$ method significantly reduces the variance of the estimates and provides more accurate estimates in regions with high likelihood values. However, in regions where $p_\varepsilon(y\mid\theta)$ approach zero, substantial deviations are observed because the logarithmic function amplifies the errors in the Monte Carlo estimates. By employing the CRNs technique, fitting a Gaussian density can significantly enhance both the accuracy and variance of the estimates.

\subsection{Three synthetic probability densities}
\begin{figure}[H]
    \centering
    \includegraphics[width=0.32\linewidth]{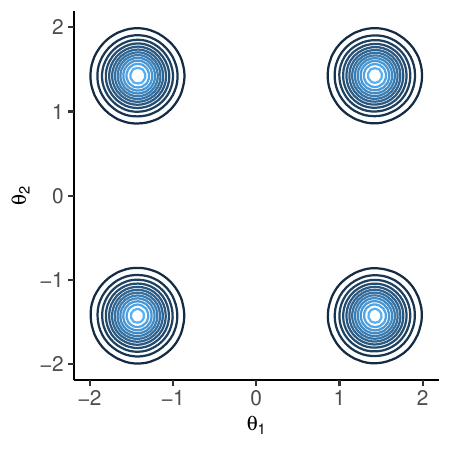}
    \includegraphics[width=0.32\linewidth]{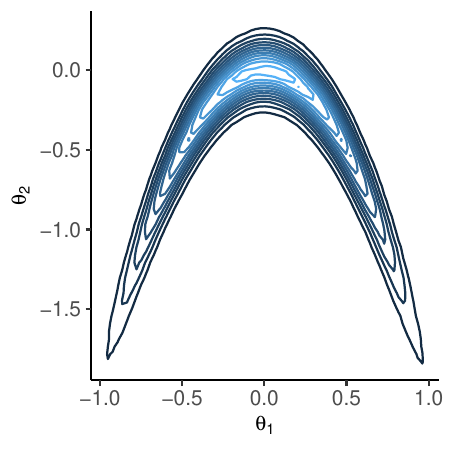}
    \includegraphics[width=0.32\linewidth]{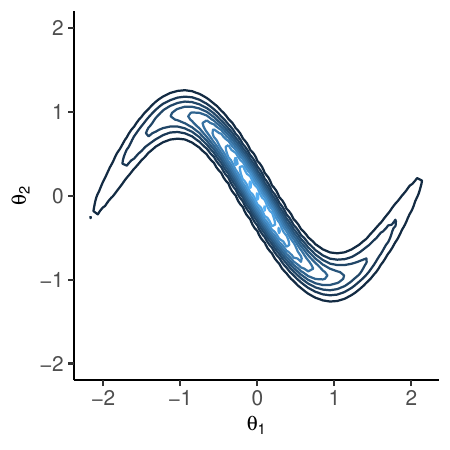}
    \caption{Posterior probability density illustration for three synthetic test functions. Labels of the test functions from left to right: Mixture Gaussian, Moon and Wave.}
    \label{fig:three synthetic}
\end{figure}

We illustrate the advantages of our approach through three synthetic examples, such as Mixture Gaussian, Moon and Wave shaped posterior. The visualization of  posterior densities are shown in Figure \ref{fig:three synthetic}. In this simulation, MCMC with a local proposal means the candidate particle $\theta^*\sim N(\theta_{t-1},\sigma^2)$ at the $t$-th iteration, and MCMC with global proposal (MCMCg) means the candidate particle $\theta^*\sim q(\cdot)$, where $q(\cdot)$ is independent with $\theta_{t-1}$, at the $t$-th iteration. The true ABC posterior is estimated using $5\times10^5$ samples, which is obtained by running importance sampling with $10^8$ samples from the prior. We use the \textit{bkde2D} function in the R package \textit{KernelSmooth} \citep{wand1994fast,wand1994kernel} to estimate the posterior density based on the posterior samples. The accuracy of different methods is measured by the KL divergence between the estimated posterior and the true posterior, which is calculated as 
$$D_{KL}(\pi_\varepsilon\mid\hat{\pi}_\varepsilon)=\frac{1}{|\Theta_{ref}|}\sum_{\thetabold\in\Theta_{ref}}\pi_\varepsilon(\thetabold\mid \ybold)\log\frac{\pi_\varepsilon(\thetabold\mid \ybold)}{\hat{\pi}_\varepsilon(\thetabold\mid \ybold)},$$ where $\pi_\varepsilon$ is the true ABC posterior density and $\hat{\pi}$ is the estimated ABC posterior density. In Figure \ref{fig:three synthetic}, $\Theta_{ref}$ is generated in a two-dimensional grid of $501^2$ points on the posterior region, that is $[-3,3]\times[-3,3]$ of Mixture, $[-2,2]\times[-5,1]$ of Moon, and $[-1,1]\times[-4,4]$ of Wave.
More details of each simulation model are provided in the Appendix C.2 of the Supplementary Material.

\begin{figure}[H]
    \centering
    \includegraphics[width=0.32\linewidth]{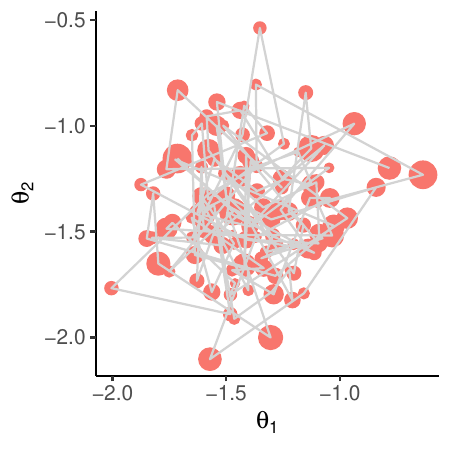}
    \includegraphics[width=0.32\linewidth]{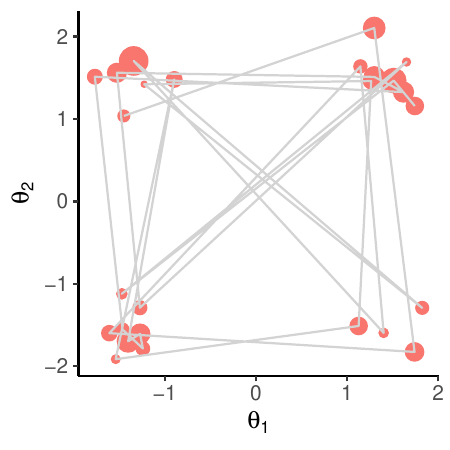}
    \includegraphics[width=0.32\linewidth]{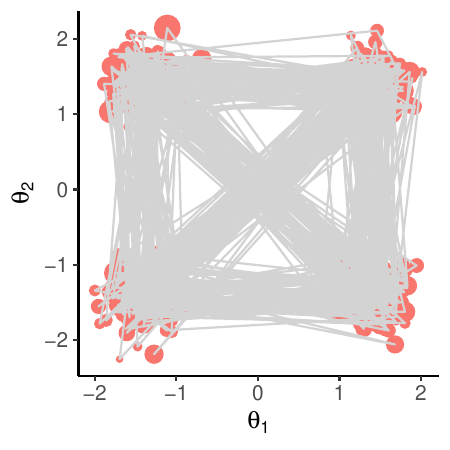}
    \caption{Trace plots of three ABC-MCMC methods for the mixture of Gaussian example: ABC-MCMC with a local proposal (left), ABC-MCMC with the prior distribution as global proposal (middle) and ABC-i-SIR with the prior distribution as the poposal and the batch size is 51 (right). The red dots and their size represent location and number of particles, the grey lines depict the movement trajectories.}
    \label{fig:three synthetic traceplot}
\end{figure}

 Firstly, we utilize the mixture Gaussian example to illustrate how ABC-i-SIR can improve the mixing of ABC-MCMC with local proposals, which fails to switch effectively between peaks. Figure \ref{fig:three synthetic traceplot} displays part of the trace plots (\ie, iteration $30,001\sim 40,000$) for three MCMC methods: ABC-MCMC with a Gaussian proposal (left), ABC-MCMC with prior distribution as the global proposal (middle), and ABC-i-SIR with prior distribution as the importance proposal with batch size $51$ (right).  The analysis reveals that ABC-MCMC with local proposals struggle to transit efficiently between peaks, while ABC-MCMC samplers with global proposals can traverse between peaks, their efficiency in exploring the local posterior region is notably low. In contrast, the ABC-i-SIR algorithm demonstrates that its particles are capable of both traversing between peaks and effectively exploring the posterior region. 

Additionally, we employ the mixture Gaussian example to investigate the influence of various proposal distributions and batch size on the performance of ABC-i-SIR. Here, we use three different distributions as the global proposal of MCMC and the importance proposal of ABC-i-SIR respectively, that is, Uniform: ${U}(-4,4)^2$, Prior: $N(0,I_2)$ and Optimal: $(N((1.425,1.425)^\intercal,0.28^2I)+N((-1.425,1.425)^\intercal,0.28^2I)+N((1.425,-1.425)^\intercal,0.28^2I)+N((-1.425,-1.425)^\intercal,0.28^2I))/4$. 
\begin{figure}[H]
    \centering
    \includegraphics[width=1\linewidth]{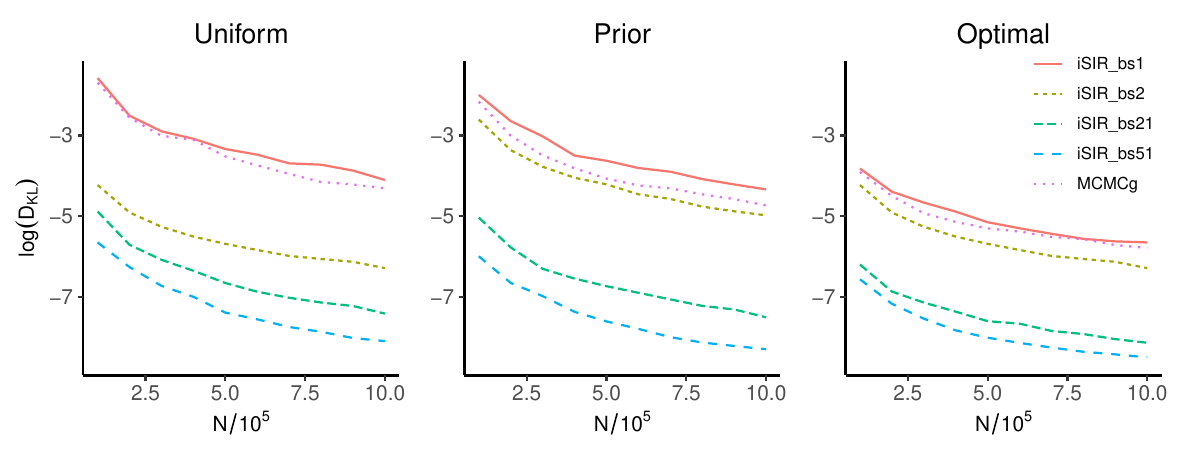}
    \caption{Comparison of ABC-i-SIR and MCMC with different proposals and batch sizes for Mixture Gaussian.  ``Optimal": a proposal close to the ABC posterior, ``Uniform": a uniform proposal, ``Prior" : the prior distribution as proposal. }
    \label{fig:Toy iSIR Proposal}
\end{figure}

Figure \ref{fig:Toy iSIR Proposal} shows that the accuracy of the ABC-i-SIR algorithm increases with the batch size. The selection of proposal distributions impacts the convergence speed of the ABC-i-SIR algorithm, but this effect gradually decreases as the batch size increases. 
This indicates that we can improve the performance of the algorithm by increasing the batch size when it is difficult to construct a good proposal distribution. This could be highly advantageous for parallel simulation models. It is also worth noting that ABC-MCMC with global proposals outperforms ABC-i-SIR with batch size 1. 

\begin{figure}[H]
    \centering
    \includegraphics[width=1\linewidth]{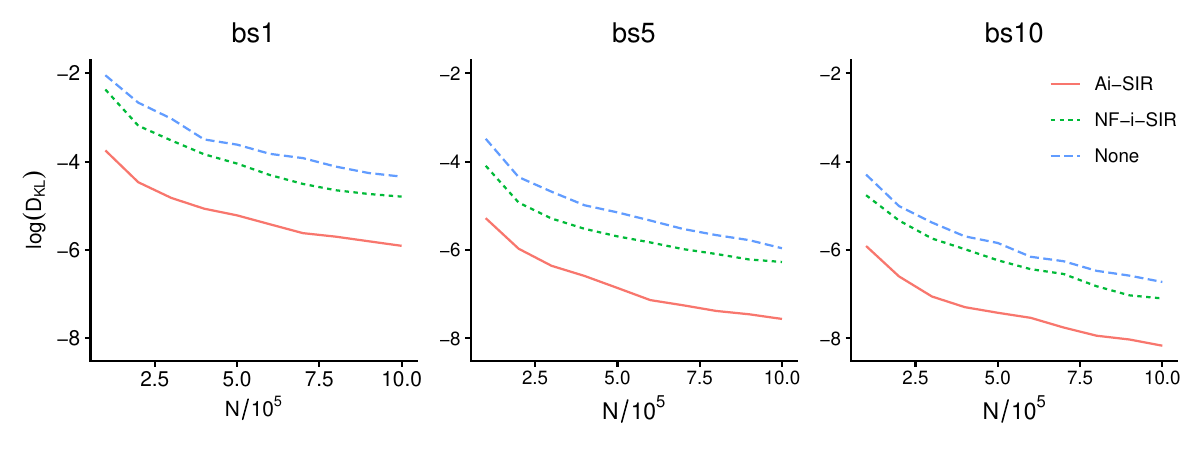}
    \caption{Comparison of different methods regarding the use of global proposal updates for the Mixture Gaussian model. ‘bs\(i\)’ denotes ABC-i-SIR with batch size \(i\). Green dashed lines: methods employing NF; red solid line: the proposed ABC-Ai-SIR algorithm; blue long-dashed lines: methods without global proposal updates.}
    \label{fig:Toy AISIR-Mix}
\end{figure}

Figure \ref{fig:Toy AISIR-Mix} presents a comparison of inference results for the Mixture Gaussian model, obtained using the Global kernel with and without adaptive updates of the importance proposal distribution. The performance of ABC-i-SIR with various batch sizes is also evaluated. Two strategies for adaptively updating the importance proposal distribution are considered: the ABC-Ai-SIR algorithm proposed in this work, and an ABC extension of \cite{samsonov2022local-global} (see Appendix Section A.1 for details). The initial proposal distribution is set as the prior, and the importance proposal distribution is updated by collecting every $2,000$ simulated data points. In the ABC-Ai-SIR algorithm, the auxiliary threshold reduction factor is set to \(\alpha = 0.8\), the target auxiliary threshold is \(\varepsilon^*_T = 5 \varepsilon\), and the normalizing flow (NF) model is updated 50 times.

\begin{figure}[H]
    \centering
    \includegraphics[width=0.24\linewidth]{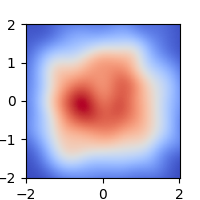}
    \includegraphics[width=0.24\linewidth]{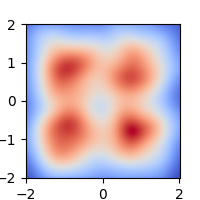}
    \includegraphics[width=0.24\linewidth]{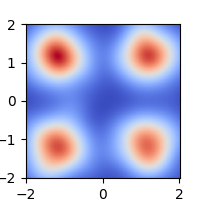}
    \includegraphics[width=0.24\linewidth]{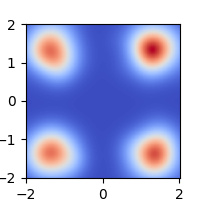}
    \caption{Evolution of the importance proposal distribution in the ABC-Ai-SIR algorithm at the 1st, 6th, 11th, and 16th updates, shown from left to right.}
    \label{fig:Toy AISIR-Mix-prop-upddate}
\end{figure}

As shown in Figure \ref{fig:Toy AISIR-Mix}, effective proposal updates can significantly improve inference results, and our proposed method demonstrates superior efficiency. Our proposed method only takes three to four minutes, whereas the NF method takes approximately half an hour. All experiments were conducted on a Mac mini equipped with an Apple M4 processor and 16 GB of RAM. Figure \ref{fig:Toy AISIR-Mix-prop-upddate} illustrates the evolution of the proposal distributions of ABC-Ai-SIR. It can be observed that the proposal distribution is steadily updated towards the target distribution over iterations.  Furthermore, the NF method directly uses the ABC posterior distribution as the update target. Due to the significant difference between the initial and target distributions, insufficient training data may lead to limited improvement in the proposed distribution, thus affecting inference performance. Further discussions on this issue, encompassing the evolution of proposal constructed by NF and detailed results concerning the Moon and Wave models, are provided in the Appendix C.3.

\begin{table}[H]
    \centering 
    \caption{The KL divergence between the estimated posterior of different ABC-MCMC methods and the ground truth for Moon and Banana data sets. $\gamma$ denotes global frequency and `bs$_i$' denotes batch size $i$.}
    \label{tab:Toy}
    \begin{tabular}{cccccccccc}
    \hline 
         & &\multicolumn{2}{c}{Mixture $({10^{-3}(10^{-4})})$}&  \multicolumn{2}{c}{Moon ($10^{-3}(10^{-4})$)} &\multicolumn{2}{c}{Wave ($10^{-5}(10^{-6})$)} \\\hline
         \multicolumn{2}{c}{MCMC}&\multicolumn{2}{c}{1658.5(1145.9)}&\multicolumn{2}{c}{13.1(85.4)}&\multicolumn{2}{c}{63.1(470.7)} \\
         \multicolumn{2}{c}{MALA}&\multicolumn{2}{c}{1650.0(922.0)}&\multicolumn{2}{c}{5.9 (33.3)}&\multicolumn{2}{c}{18.3(86.6)}\\\hdashline
         &$\gamma$&bs$_5$&bs$_{10}$&bs$_5$&bs$_{10}$&bs$_5$&bs$_{10}$\\
         \multirow{5}*{\rotatebox{90}{GL-MCMC}}
         &0.2&2.4(2.3)&1.9(2.0)&1.9(9.1)&1.3(5.1)&12.1(26.3)&9.7(14.5)\\
         &0.4&\textbf{2.1}(2.2)&1.6(1.2)&1.4(4.9)&0.7(2.1)&11.3(24.7)&6.7(6.3)\\
         &0.6&2.2(2.1)&1.4(1.4)&1.2(2.9)&0.7(1.9)&8.4(9.0)&5.2(3.5)\\
         &0.8&2.3(1.8)&1.3(1.0)&\textbf{0.9}(3.1)&0.6(1.9)&7.2(8.0)&4.1(4.1)\\
         &1.0&2.4(1.6)&\textbf{1.2}(0.7)&1.1(2.2)&\textbf{0.4}(0.7)&\textbf{6.5}(5.9)&\textbf{3.6}(3.6)\\\hline
    \end{tabular}
\end{table}
Tables \ref{tab:Toy}  presents a comparison of several ABC-MCMC methods for the Mixture Gaussian Moon, and Wave.  GL-MCMCs are the combinations of MALA and ABC-i-SIR with a prior distribution as the importance proposal. The step size of MALA is chosen by the sequential optimal algorithm based on ESJD. Notably, GL-MCMC degenerates into MALA when $\gamma=0$ and ABC-i-SIR when $\gamma=1.0$. Here, we repeated the MCMC algorithm  10 times, each with 1,000,000 iterations. Table \ref{tab:Toy} shows the KL divergences between the estimated posterior of different ABC-MCMC methods and the ground truth and their standard deviations. In the three examples mentioned, the combination of global proposal and local proposal has shown superior performance compared to ordinary ABC-MCMC and MALA algorithms, especially in the case of the multi-modal Mixture Gaussian example. Compared with the MCMC algorithm, the MALA algorithm, which utilizes the gradient of the log-likelihood function, has a more accurate posterior estimate and a larger effective sample size (\ie~Table \ref{tab:Toy ESS}), especially for models where the posterior distribution exhibits strong correlations, such as Moon and Wave. 
When using a batch size of 5, the GL-MCMC algorithm with \(\gamma=0.4\) in the Mixture Gaussian example and \(\gamma=0.8\) in the Moon example performs the best. This suggests that the combination of global and local proposals may be particularly effective when the global exploration ability of the algorithm is relatively weak.

\begin{table}[H]
    \centering 
    \caption{Effective sample size of the posterior samples obtained by MCMC and MALA.}
    \label{tab:Toy ESS}
    \begin{tabular}{ccccccc}
    \hline 
          &\multicolumn{2}{c}{Mixture}&  \multicolumn{2}{c}{Moon} &\multicolumn{2}{c}{Wave} \\\hline&$\theta_1$&$\theta_2$&$\theta_1$&$\theta_2$&$\theta_1$&$\theta_2$\\\hdashline
         MCMC&4050&3940&150&164&451&956\\ 
         MALA&7589&7788&962&691&1474&2710\\\hline
    \end{tabular}
\end{table}

\section{Real Examples}
\label{sec:real}

\subsection{Example 1: Roman binary microlensing events}
A binary-lens, single-source (2L1S) microlensing event is an astronomical phenomenon where the light from a distant source star is bent and magnified as it passes through a binary star system. This phenomenon offers a unique opportunity to study and discover exoplanets. Fast and automated inference of 2L1S microlensing events using Markov Chain Monte Carlo (MCMC) methods faces two main challenges: (a). the high computational cost associated with likelihood evaluations using microlensing simulation codes, and (b). a complex parameter space characterized by a negative-log-likelihood surface with numerous narrow and deep local minima.

In this section, we study the 2L1S microlensing model investigated in \cite{zhang2021real}. We fix the angle of the source trajectory relative to the projected binary lens axis at $\alpha=110$ degrees and the time of closest approach at $t_0=60.0$ day. There are in total five unknown parameters in the model: binary lens separation ($s$), mass ratio ($q$), impact parameter ($u_0$), Einstein ring crossing timescale ($t_E$), and source flux fraction ($f_s$).

The 2L1S magnification sequences are simulated using the microlensing code \textit{MulensModel} \citep{poleski2019modeling}, over a span of 144 days with a cadence of 15 minutes, the number of observations $N_{\ybold}=13825$. The observed data $\ybold$ is generated using the parameters $(s,q,u_0,t_E,f_s) = (10^{-0.2},10^{-2.5},0.2,10^{1.6},0.2)$. The discrepancy between the observed data $\ybold$ and the simulated data $\xbold$ is defined as $\Delta(\xbold,\ybold)=\frac{1}{N_{\ybold}}\sum_{i=1}^{N_{\ybold}}|\ybold_i-\xbold_i|$. The kernel function employed is a Gaussian kernel with a bandwidth $\varepsilon=0.003$. We then simulate 2L1S events using the analytical priors: 
\begin{equation}
    \begin{aligned}
        &s\sim\text{LogUniform}(0.2,5), ~q\sim\text{LogUniform}(10^{-6},1),~ u_0\sim\text{Uniform}(0,2),\\
        &t_E\sim\text{TruncLogNorm}(1,100,\mu=10^{1.15},\sigma=10^{0.45}),~f_s\sim\text{LogUniform}(0.1,1).
    \end{aligned}
    \label{eq: mulen prior}
\end{equation} In all schemes that employ a local proposal, the local proposal is set to 
\begin{equation}
    \begin{aligned}
        &\log(s^*)\sim N(\log(s),0.2^2),~\log(q^*)\sim N(\log(q),0.2^2),~u_0^*\sim N(u_0,0.01^2),\\
        &\log(t_E^*)\sim N(\log(t_E),0.05^2), ~\log(f_s^*)\sim N(\log(f_s),0.05^2).
    \end{aligned}
    \label{eq: mulenlocal_prop}
\end{equation}

\begin{figure}[H]
        \centering
        \includegraphics[width=1\linewidth]{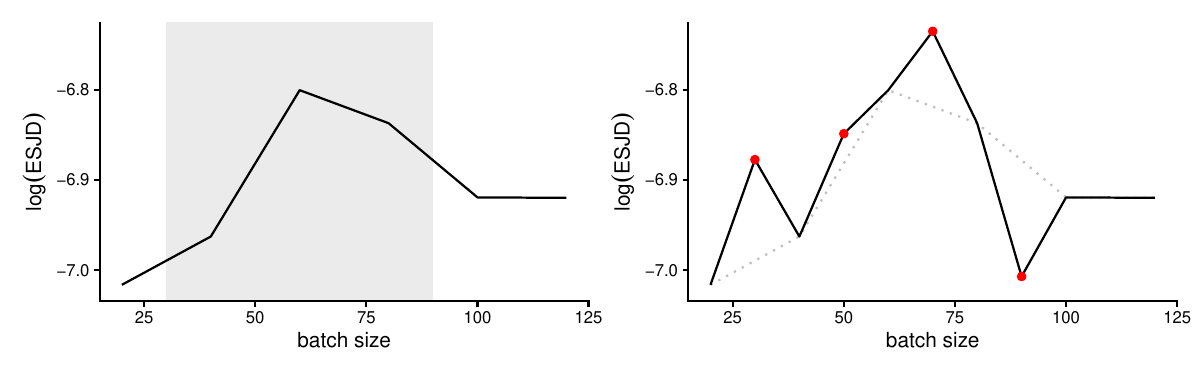}
        \caption{Sequential optimization of hyper-parameter $N_b$ with constrained condition $C=5$.}
        \label{fig:ESJD}
\end{figure}

First, we select the hyperparameters for our GL-ABC-MCMC method. We employ ABC-i-SIR with the prior distribution (\ie, Eq. \eqref{eq: mulen prior}) as the importance proposal. To manage the simulation cost, we fix the average number of simulations per iteration to \(C=5\), which constrains the global frequency \(\gamma\) and the batch size \(N_b\) via \(\gamma N_b + (1-\gamma) = C\). For each candidate hyperparameter configuration, GL-ABC-MCMC is run for 5,000 iterations, and the entire procedure is repeated five times to assess variability. Figure \ref{fig:ESJD} illustrates the selection of \(N_b\) through sequential optimization, which leads us to set \(N_b = 70\). From this point onward, for all GL-ABC-MCMC runs in this example, $N_b$ is fixed at 70 unless otherwise noted.

\begin{figure}[H]
    \centering
    \includegraphics[width=0.32\linewidth]{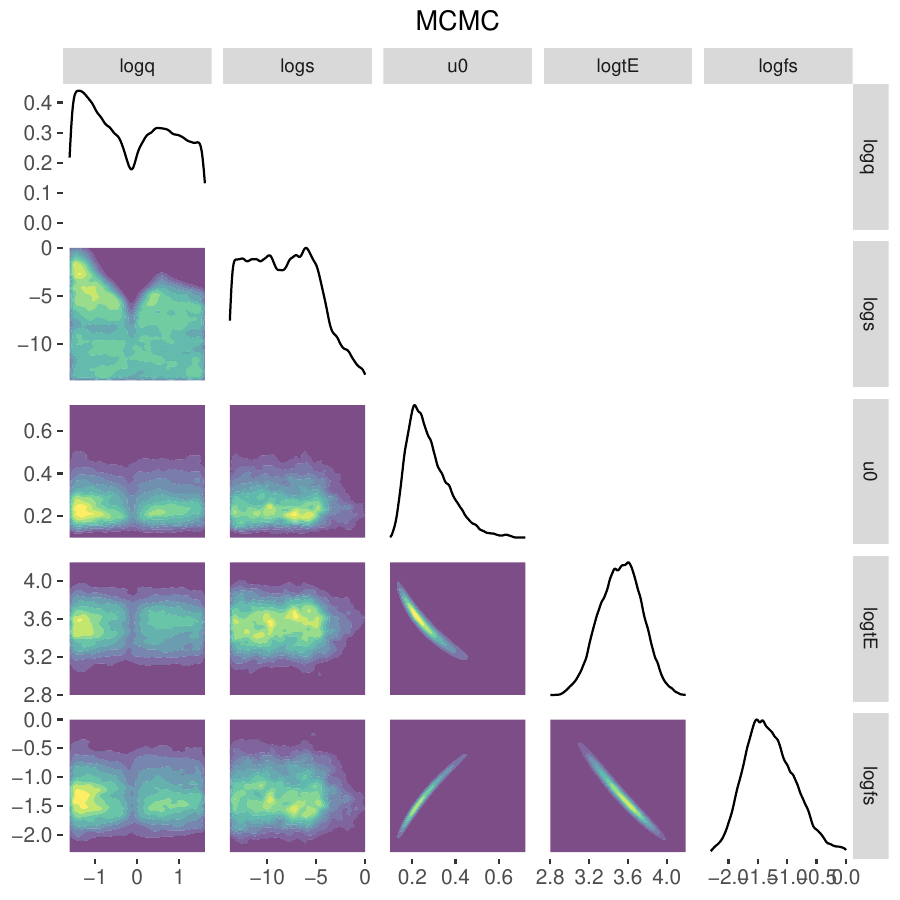}
    \includegraphics[width=0.32\linewidth]{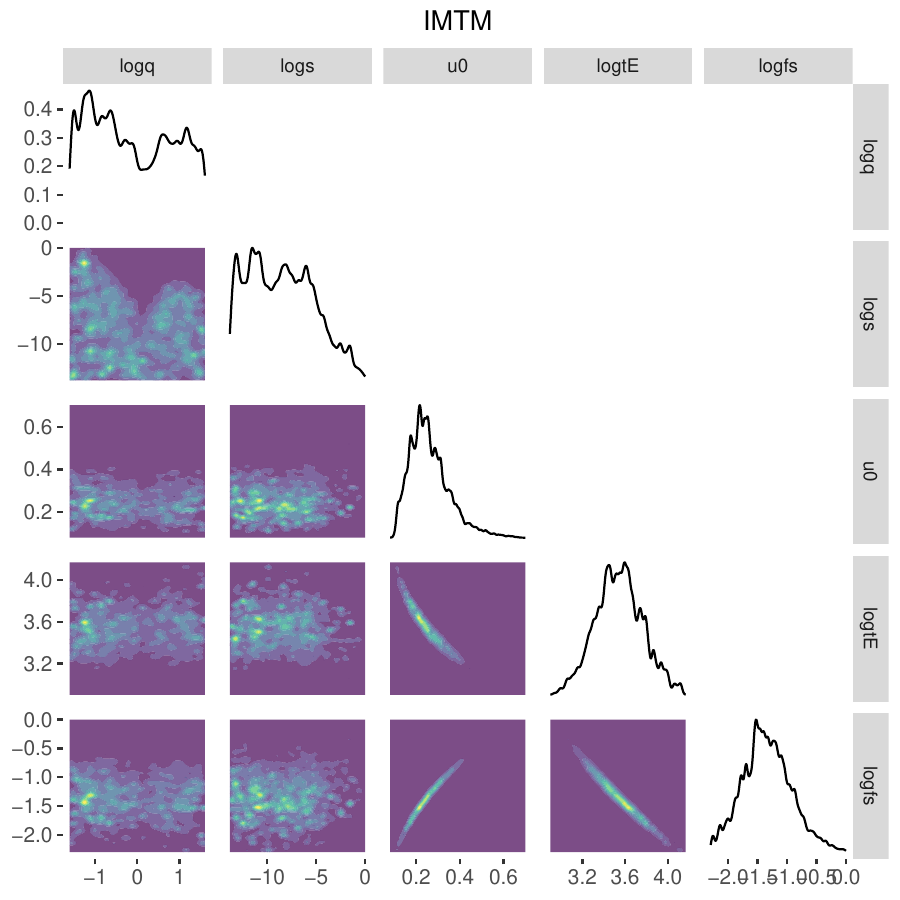}
    \includegraphics[width=0.32\linewidth]{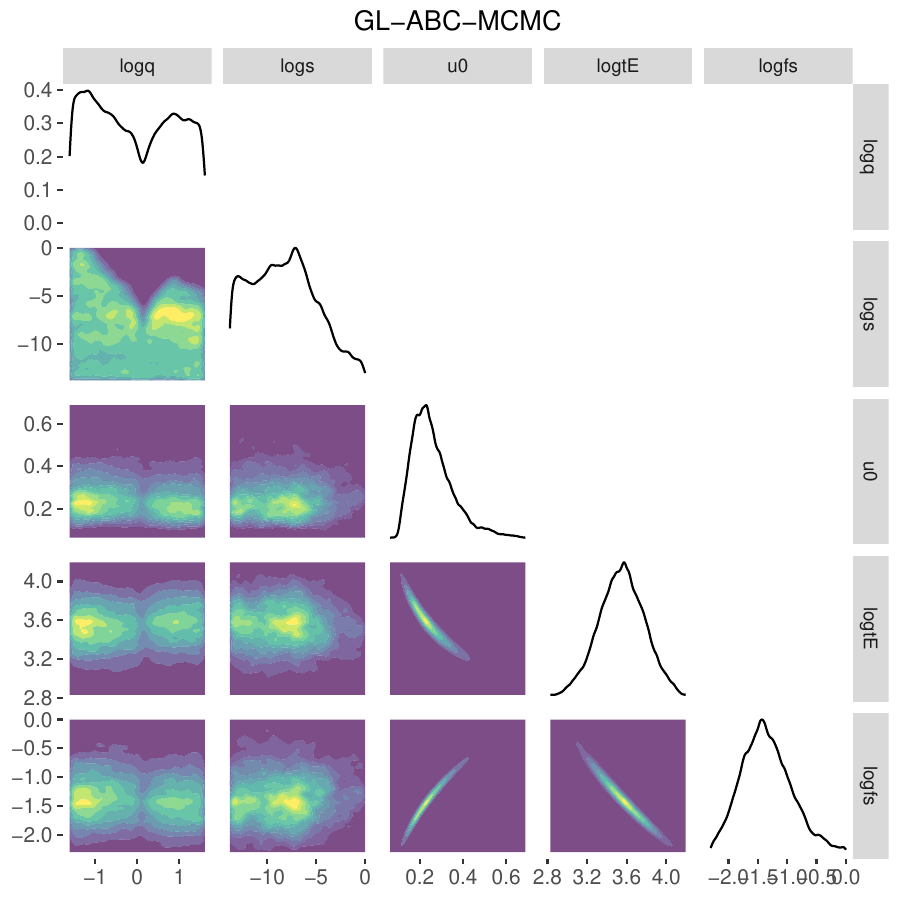}
    \includegraphics[width=0.32\linewidth]{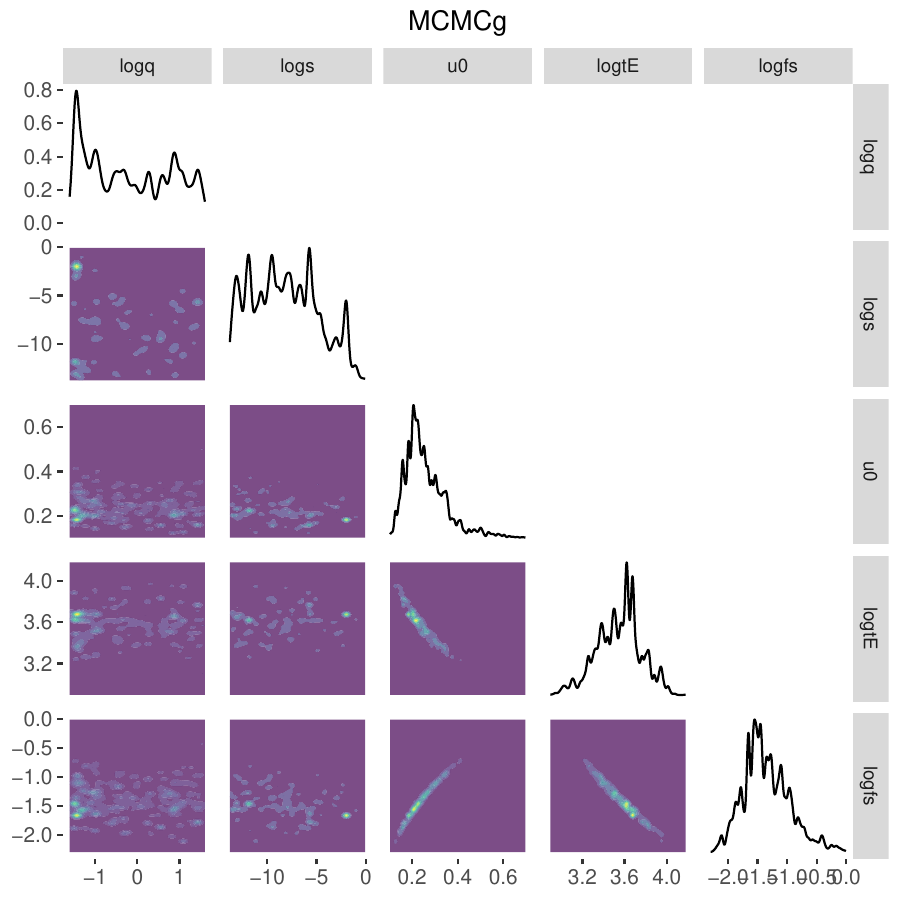}
    \includegraphics[width=0.32\linewidth]{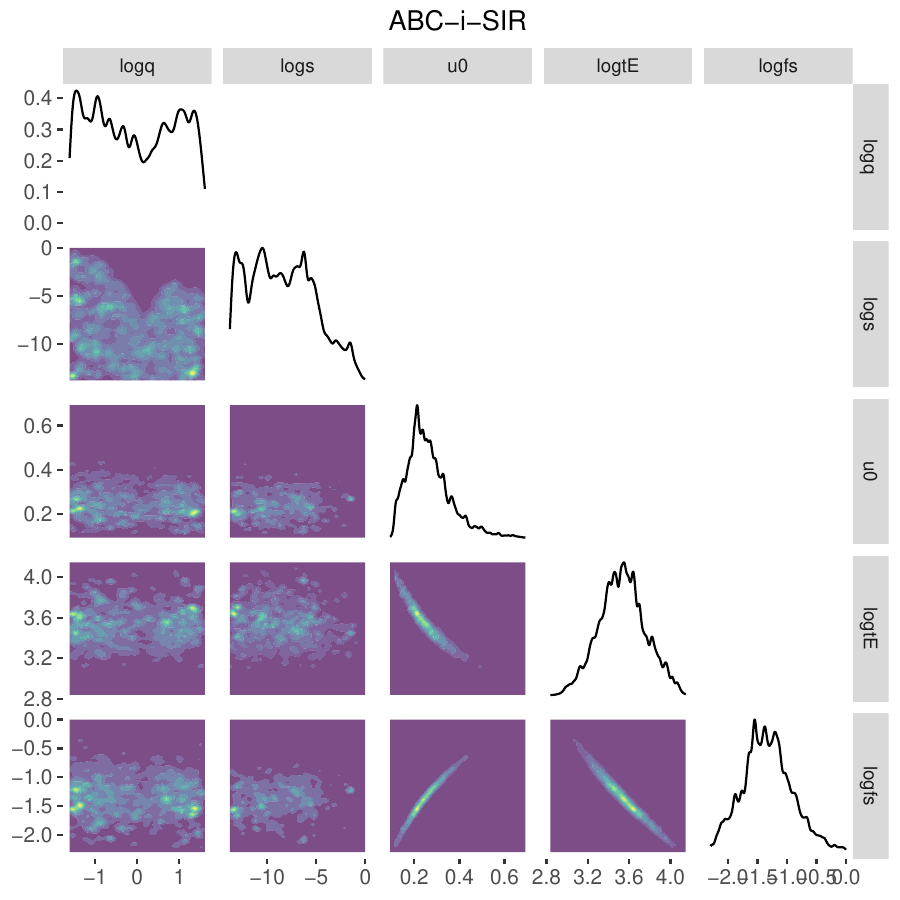}
    \includegraphics[width=0.32\linewidth]{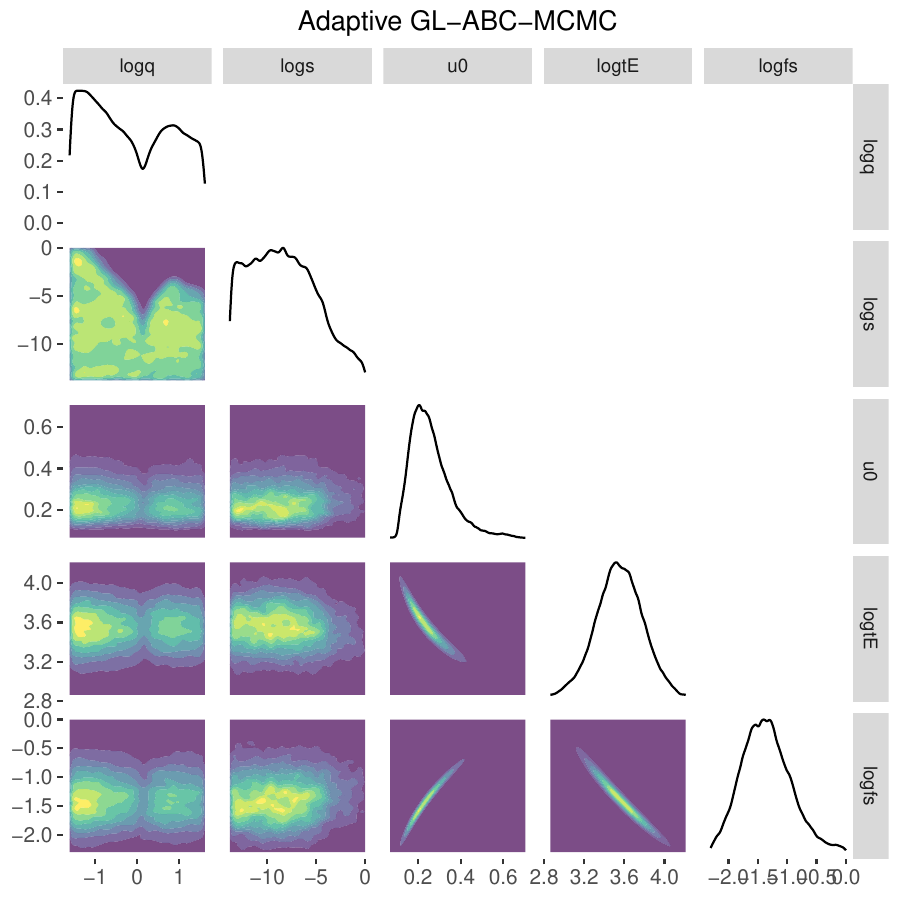}
    
    \caption{Visualization of the posterior distributions estimated by various ABC-MCMC methods.}
    \label{fig:mulen_2D}
\end{figure}

Figure \ref{fig:mulen_2D} presents the posterior estimates obtained via six MCMC methods with over 500,000 iterations:
\begin{itemize}
    \item \textbf{MCMC}: Basic ABC-MCMC utilizing the proposal defined in Eq. \eqref{eq: mulenlocal_prop}.
    \item \textbf{MCMCg}: ABC-MCMC with the proposal given by Eq. \eqref{eq: mulen op}.
    \item \textbf{ABC-i-SIR}: Importance sampling with $N_b=5$ and proposal distribution specified by Eq. \eqref{eq: mulen op}.
    \item \textbf{IMIM}: Independent Multiple Try Metropolis \citep{MARTINO2018134}, employing the same sets as ABC-i-SIR.
    \item \textbf{GL-ABC-MCMC}: Global-local ABC-MCMC where the global move corresponds to ABC-i-SIR with $N_b=70$, using Eq. \eqref{eq: mulen prior} as the importance proposal distribution. The probability of a global move, $\gamma$, satisfies $\gamma N_b + (1 - \gamma) = 5$.
    \item \textbf{Adaptive GL-ABC-MCMC}: Extends GL-ABC-MCMC by adaptively updating the proposal according to Algorithm \ref{alg:Ai-SIR}, with auxiliary target threshold $\varepsilon_T^* = 0.015$, threshold decay parameter $\alpha=0.6$, and the importance proposal updated every $S=50$ global proposals.
\end{itemize}

The substantial discrepancy between prior and posterior leads to negligible acceptance rates when the prior is used as the global proposal, as observed in MCMCg, IMIM, and ABC-i-SIR. To address this, we construct an efficient proposal distribution
\begin{equation}
    q(\thetabold) = \mathcal{N}(\thetabold; \boldsymbol{\mu}, \Sigma),
    \label{eq: mulen op}
\end{equation}
where $\mathcal{N}(\thetabold; \boldsymbol{\mu}, \Sigma)$ denotes a Gaussian distribution with mean vector $\boldsymbol{\mu}$ and covariance matrix $\Sigma$. The parameters $\boldsymbol{\mu}$ and $\Sigma$ are estimated from the 500 realizations with the smallest discrepancies among 5,000 prior draws.

As illustrated in Figure \ref{fig:mulen_2D}, ABC-MCMC with a global proposal, ABC-i-SIR, and IMIM exhibit random-walk behavior, hindering efficient exploration. In contrast, the standard ABC-MCMC, GL-ABC-MCMC, and adaptive GL-ABC-MCMC successfully explore both modes of the posterior distribution. 

\begin{table}[H]
        \centering
        \caption{Comparison of effective sample sizes of different methods: ABC-MCMC, GL-ABC-MCMC, GL-ABC-MCMC$_c$ (using Eq. \eqref{eq: mulen op} as the importance proposal), Adptive GL-ABC-MCMC. The average number of simulations per iteration of GL-ABC-MCMC is $5$.}
        \begin{tabular}{cccccc}
             \hline
             Method&$\log(s)$&$\log(q)$&$u_0$&$log(t_E)$&$log(f_s)$\\\hline
            ABC-MCMC& 1052.428 & 132.8922 & 206.6277 & 271.4922 & 228.3610\\
            GL-ABC-MCMC& 1133.471 & 314.4186 & 481.5078 & 476.1921 & 464.2407\\
            GL-ABC-MCMC$_c$& 2252.269 & 1183.1576 &1863.4926 & 1649.8262 &1647.7599\\
            Aaptive GL-ABC-MCMC & 7516.687 & 6630.4125 & 8273.9112 &7362.3383 &7501.7451
             \\\hline
        \end{tabular}
        \label{tab:ESS}
    \end{table}
    \begin{figure}[H]
    \centering
    \includegraphics[width=1\linewidth]{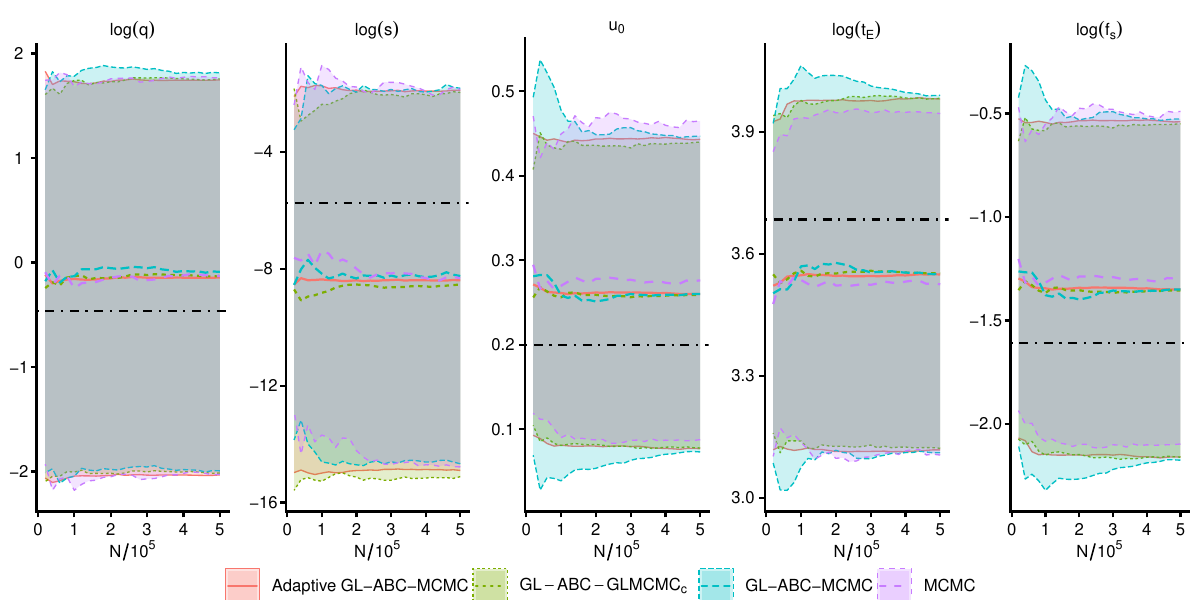}
    \caption{Posterior mean trajectories and 95\% credible intervals versus iteration number for various ABC-MCMC methods. The black dotdash lines are true values we used to simulate the data.}
    \label{fig: mulen_mean}
\end{figure}

Table \ref{tab:ESS} compares the effective sample sizes (ESS) across the standard ABC-MCMC, GL-ABC-MCMC, GL-ABC-MCMC$_c$ and adaptive GL-ABC-MCMC approaches. Figure \ref{fig: mulen_mean} examines the evolution of the estimated posterior means as a function of MCMC iterations. 
The GL-ABC-MCMC$_c$ is constructed with importance proposal shown in Eq.~\eqref{eq: mulen op}. 
The results show that GL-ABC-MCMC strategies outperform standard ABC-MCMC that relies on local proposals in terms of both ESS and convergence speed. For this complex model, a global move constructed from the prior offers limited improvement and falls short of the performance achieved by the constructed proposal distribution. The adaptive method we proposed exhibits substantially superior performance: with the same number of iterations, the ESS is tens of times larger than that of ABC-MCMC, and the posterior means stabilize rapidly. 

We attempted to use normalizing flows to update the proposal distribution under the same parameter settings as the GL-ABC-MCMC algorithm. Since the posterior region is much narrower compared to the prior, only about one hundred samples with non-zero weights were obtained after $50$ global moves (\ie, $3500$ samples in total). The limited number of effective training samples caused issues in training the normalizing flow, resulting in the support of the proposal distribution deviating from the posterior region. Consequently, the global moves could not perform effectively. Therefore, the results of NF method is not used for comparison. Nevertheless, in the Appendix Figure C.4, we present the evolution of the proposal distributions using both the method proposed in this work and the NF method. During the training of the NF, if 50 global training iterations pass without obtaining 500 training samples with importance weights greater than zero, the training is paused until 500 such samples are collected. The importance proposals constructed by the NF method fail to fully cover the posterior region, further highlighting the necessity and effectiveness of introducing intermediate target distributions when dealing with complex posterior models. More details of this experiment are described in Appendix C.4.

\subsection{Example 2: A stochastic differential equation example}

We consider parameter estimation of a high dimensional non-linear example, the Van der Pol oscillator model \citep{kandepu2008applying,sarkka2015posterior}. The model is described by the following second-order non-linear ODE:
$$
\frac{{d^2x(t)}}{{dt^2}} - \mu(1 - \epsilon x^2(t))\frac{{dx(t)}}{{dt}} + x(t) = f(t).
$$
The unknown parameters in the model are $\mu$ and $\epsilon$, and $f(t)$ represents an additional unknown forcing term.
To model the forcing term $f(t)$, we assume it is a combination of white noise and a stochastic resonator $c(t)$, which is formed by the sum of $N$ harmonic components $c_n(t)$:
$$
\frac{{d^2c_n(t)}}{{dt^2}} = -(n\omega_c)^2c_n(t) + \sigma_n\epsilon_n(t).
$$
Here $\omega_c$ represents the angular velocity of the force process, $\sigma_n$ represents the strength of the noise process driving the $n$th harmonic, and $\epsilon_n(t)$ represents a white noise process. In this section, we set $N=2$. For simplicity, we assume that $\sigma_n = \sigma_c$ for all $n=1,2$. Consequently, the parameter vector $\boldsymbol{\theta} = (\epsilon, \mu, \sigma, \omega_c, \sigma_c)$ is five-dimensional, and the state variable $\xbold = (x, \dot{x}, c_1, \dot{c}_1, c_2, \dot{c}_2)$ has $6$ components. The full SDE model can be represented as:
$$
\begin{aligned}
dx(t) &= \dot{x}(t)dt, \\
d\dot{x}(t) &= \mu(1-\epsilon x^2(t))\dot{x}(t)dt - x(t)dt + (c_1(t)+c_2(t))dt, \\
dc_n(t) &= \dot{c}_n(t)dt, \\
d\dot{c}_n(t) &= -(\omega_c)^2c_n(t)dt + \sigma_c dW_n(t), \quad \text{for } n=1,2.
\end{aligned}
$$
Here $dW_n(t)$ represents the differential of a standard Wiener process. The measurements are obtained by adding noise to the state of the Van der Pol oscillator:
$$
y_k = x(t_k) + r_k, \quad r_k \sim \mathcal{N}(0, \sigma^2).
$$
The observed data $\ybold$ was simulated from the system using the parameter values $\boldsymbol{\theta} = (1, 1/2, 1/10,$ $ \pi/5, 1/100)$ over the time interval $t \in [0, 40]$, with a sampling period of $\Delta t = 1$. Since $\thetabold\ge 0$, we take Gamma distributions as  prior distributions $\pi(\thetabold)$: $\epsilon\sim Gamma(5,1)$, $\mu\sim Gamma(3,5)$, $\sigma\sim Gamma(5,15)$, $\omega_c\sim Gamma(5,10)$, $\sigma_c\sim Gamma(2,15)$. Here, we choose the kernel function $K_\varepsilon(\cdot,\ybold)=N(\cdot;\ybold,\varepsilon^2)$ with ABC threshold $\varepsilon=0.2$. 

We compare the efficiency of ABC-MCMC, ABC-i-SIR, and GL-ABC-MCMC. The importance proposal distribution of ABC-i-SIR is the prior distribution and the batch size is 20. We choose the global frequency $\gamma$ based on cESJD.
\begin{table}[H]
    \centering
    \caption{cESJD values under different global frequency parameter.}
    \begin{tabular}{ccccccc}
    \hline
    $\gamma$& 0&0.2&0.4&0.6&0.8&1.0\\\hline
   cESJD & 0.0024&    0.0060 &   0.0096 &   0.0108  &  0.0150 &   0.0204\\\hline   \end{tabular}
    
    \label{tab:SDEgf}
\end{table}

\begin{figure}[H]
    \centering
    \includegraphics[width=0.95\linewidth]{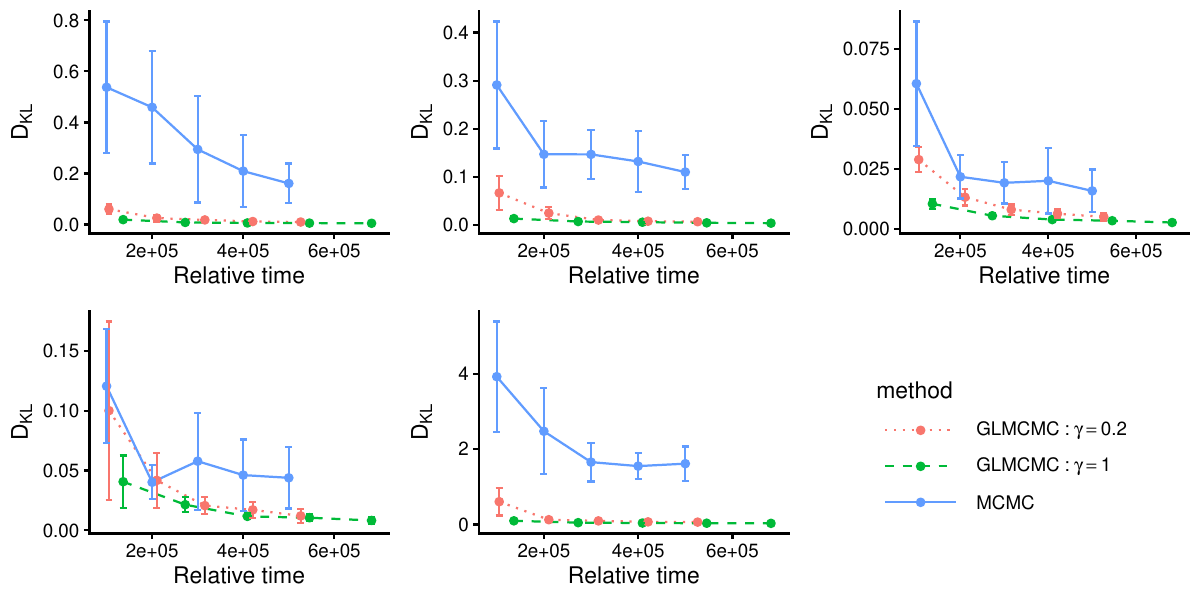}
    \caption{Comparasion of three ABC-MCMC methods (ABC-MCMC, GL-ABC-MCMC with batch size 20 and global frequency $1$ and $0.2$). Error bars denote $95 \%$ confidence intervals for $D_{KL}$ across 10 replicates. First row: $\epsilon$, $\mu$, $\sigma$; Second row: $\omega_c$, $\sigma_c$.}
    \label{fig:SDE KL}
\end{figure}

\begin{figure}[H]
    \centering
    \includegraphics[width=0.9\linewidth]{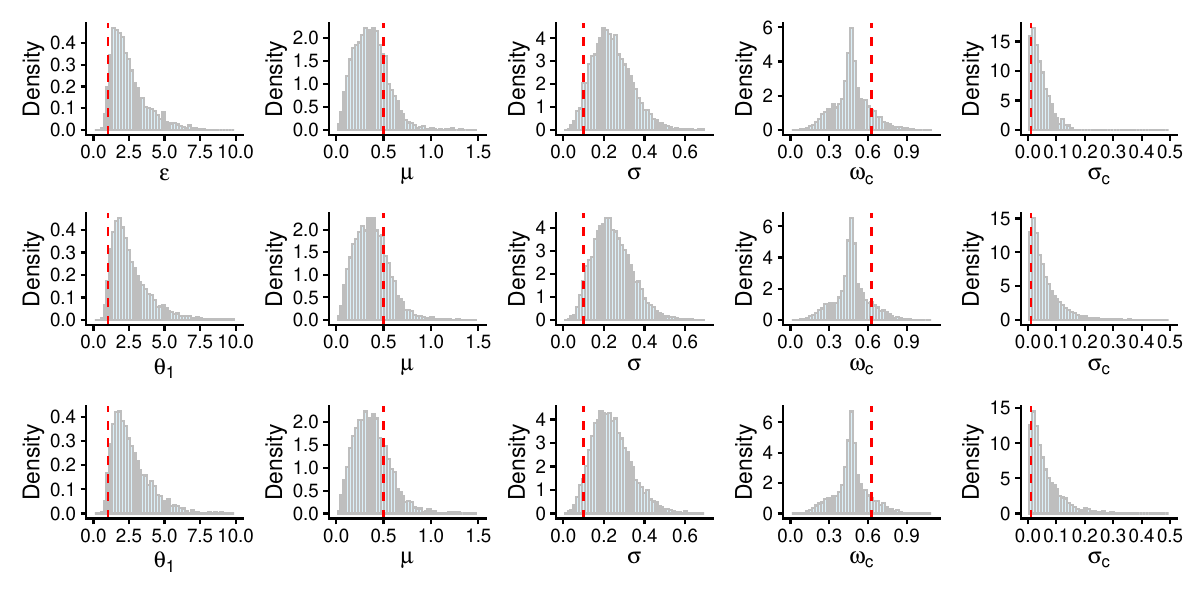}
    \caption{One-dimensional marginal distributions obtained by ABC-MCMC (first row), Global-Local ABC-MCMC with global frequency $\gamma=0.2$ (second row), and $\gamma=1$ (third row). The results correspond to 500,000 total iterations. The red line indicates the true value.}
    \label{fig:SDE-hist}
\end{figure}

\begin{figure}[H]
    \centering
    \includegraphics[width=0.9\linewidth]{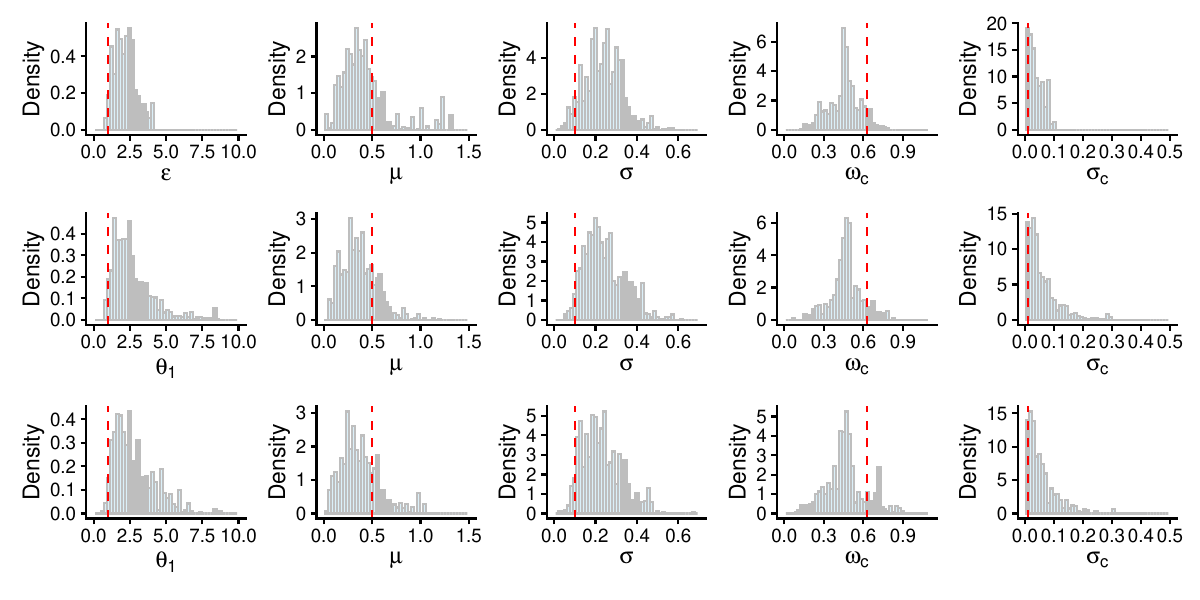}
    \caption{One-dimensional marginal distributions obtained by ABC-MCMC (first row), Global-Local ABC-MCMC with global frequency $\gamma=0.2$ (second row), and $\gamma=1$ (third row). The results correspond to the first 20,000 iterations of 500,000 total iterations. The red line indicates the true value.}
    \label{fig:SDE-hist-partial}
\end{figure}

Here, we consider the simulation time cost and choose the optimal global frequency $\gamma=1$ from candidate set shown in Table \ref{tab:SDEgf}. Additionally, we select a `worst' $\gamma=0$ and a `sub-worst' one $\gamma=0.2$ as references. We obtain the true posterior by running 5 ABC-MCMC chains with $1,000,000$ iteration, and discard the first 50,000 iterations as the burn-in. 
Here, we use \textit{``density"} function in \textit{R} to estimate the 1D marginal density with 512 points on the posterior region, that is $[0,12]\times[0,1.5]\times[0,1]\times[0,1.5]\times[0,0.5]$. And the KL divergence is calculated by
$$D_{KL}(\pi_\varepsilon(\theta_i)\mid\hat{\pi}_\varepsilon(\theta_i))=\frac{l_i}{|\Theta_{i,ref}|}\sum_{\theta_i\in\Theta_{i,ref}}\pi_\varepsilon(\theta_i\mid \ybold)\log\frac{\pi_\varepsilon(\theta_i\mid \ybold)}{\hat{\pi}_\varepsilon(\theta_i\mid \ybold)},$$ where $\pi_\varepsilon(\theta_i)$ is the true ABC posterior marginal density of $\theta_i$, $\hat{\pi}_\varepsilon(\theta_i)$ is the estimated ABC posterior  marginal density of $\theta_i$, and $l_i$ is the length of the interval for the posterior estimate. Here we calculate KL divergence of different methods at iteration numbers $10^5$, $2\cdot10^5$, $\ldots$, $10^6$. We take the time executing one MCMC iteration as the baseline. Figure \ref{fig:SDE KL} shows the KL divergence of different methods varying with time cost. The methods proposed in this article (\ie,~GL-ABC-MCMC and ABC-i-SIR) converge faster than ordinary MCMC, and admit smaller variance. Figure \ref{fig:SDE-hist} shows the one-dimensional marginal distribution of the parameters. Figure \ref{fig:SDE-hist-partial} shows the results correspond to the first 20,000 iterations out of 500,000 total iterations. These results reveal that ABC-MCMC tends to get trapped in local modes, resulting in insufficient coverage of the tails (tail deficiency). Moreover, it may spend excessive time dwelling in low-probability regions, which adversely affects sampling efficiency and the accurate characterization of the posterior distribution.


\section{Conclusion}
\label{sec:conc}

In this article, we address the problem of Bayesian inference for models with intractable likelihood functions. Markov chain Monte Carlo (MCMC) methods are frequently combined with approximate Bayesian computation (ABC) to facilitate likelihood-free inference. However, MCMC algorithms relying solely on local moves often struggle to adequately explore posterior distributions with complex geometries. To overcome this limitation, we propose a novel framework that integrates both efficient global and local moves, thereby significantly enhancing the posterior exploration capability within the likelihood-free inference context.

Specifically, we develop a likelihood-free variant of the iterated Sequential Importance Resampling (i-SIR) algorithm to serve as an efficient global move within ABC-MCMC algorithms. For high-dimensional parameter spaces, we employ an ABC adaptation of the Metropolis-adjusted Langevin algorithm (MALA) as an effective local move, where the gradient of the log-likelihood function is approximated via the common random numbers technique.

To further improve the overall efficiency of the proposed ABC-MCMC scheme, we introduce two adaptive strategies. First, kernel density estimation is utilized to progressively refine the importance proposal distribution by introducing a sequence of intermediate target distributions.
\cite{samsonov2022local-global} proposed constructing the importance proposal distribution for i-SIR using normalizing flows. Although normalizing flows are powerful for modeling complex distributions, their performance can deteriorate when the initial proposal is far from the target distribution, particularly in ABC inference where many samples receive negligible weights, leading to unstable updates and computational inefficiency. 
Second, we define a criterion based on a unit cost version of the expected squared jumping distance (ESJD) to quantitatively assess the exploration capability of the ABC-MCMC kernel. Building on this, a sequential optimization algorithm is designed to select the hyperparameters governing the Global-Local ABC-MCMC moves.

Moreover, we rigorously establish the V-geometric ergodicity of the Global-Local ABC-MCMC algorithm that employs ABC-i-SIR as the global move. This theoretical result implies that, under suitable regularity conditions, the mixing rate of the proposed Markov kernel outperforms that of ABC-MCMC methods relying exclusively on local moves. Extensive numerical experiments on both synthetic and real-world models demonstrate that our algorithm achieves superior performance compared to existing approaches. A Python package \texttt{glabcmcmc} implementing this method is available at \url{https://github.com/caofff/GL-ABC-MCMC}. We also refer readers to \cite{cao2025glabcmcmc} for more detailed information of this package.

Several avenues for future research and improvement arise from this work. First, the ABC tolerance parameter \(\epsilon\) is currently pre-specified prior to running the ABC-MCMC algorithm. Following the framework proposed by \cite{del2012adaptive}, an important extension is to integrate our Global-Local move scheme within an ABC sequential Monte Carlo (ABC-SMC) framework and explore adaptive strategies for tuning move parameters. Second, the computation of the Metropolis–Hastings acceptance probability, as well as the implementation of ABC-i-SIR, requires generating synthetic data from the simulation model,  which can be computationally expensive for complex models. Future efforts will focus on accelerating ABC-MCMC inference by leveraging approximation techniques such as those in \cite{doi:10.1080/10618600.2024.2379349, Likelihood-freeapproximateGibbssampling}. Third, the gradient estimation in approximate Bayesian methods for high-dimensional parameter spaces often requires a large number of simulations. To address this challenge, we plan to investigate variance reduction via control variates and develop methods to optimally determine simulation sizes for gradient approximation.

\section*{Acknowledge}
This work was supported by the National Natural Science Foundation of China (12131001 and 12101333), the startup fund of ShanghaiTech University, the HPC Platform
of ShanghaiTech University, the Fundamental Research Funds for the Central Universities, LPMC, and KLMDASR. The authorship is listed in alphabetic order. 

\bigskip
\begin{center}
{\large\bf SUPPLEMENTAL MATERIALS}
\end{center}

\begin{description}
\item[Appendix A:] Some details of algorithms not shown in the main text.

\item[Appendix B:] The proofs of all theoretical results presented in the main text.

\item[Appendix C:] Some numerical results and some details of setups not shown in the main text.
\end{description}

\bibliographystyle{chicago}
\bibliography{Bibliography-MM-MC}

\end{document}